\documentclass[11pt,]{article}
 \usepackage{graphicx}
 \usepackage[cp1251]{inputenc}  
\begin{document}

 \bigskip

 \bigskip

 \centerline {\bf ANALYSIS OF REGULARITY/CHAOTICITY OF THE GLOBULAR CLUSTERS}
 \centerline {\bf DYNAMICS IN THE CENTRAL REGION OF THE MILKY WAY}

  \bigskip

  \bigskip

   \centerline{\bf
            A.~T.~Bajkova\footnote [1]{bajkova@gaoran.ru} (1), A.~A.~Smirnov (1),
            V.~V.~Bobylev (1)
            }
  \bigskip
  \bigskip
 \centerline{\small \it (1)Central (Pulkovo) Astronomical Observatory RAS, St. Petersburg, 196140 Russia}
  \bigskip

  \bigskip

{{\bf Abstract.}
We analyzed the regularity/chaoticity of the orbits of 45 globular clusters in the central region of the Galaxy with a radius of 3.5 kpc, which are subject to the greatest influence from an elongated rotating bar. Various analysis methods were used, namely, methods for calculating the Highest Lyapunov Exponents, the method of Poincar$\acute{e}$ sections, the frequency method based on the calculation of fundamental frequencies, as well as the visual assessment method. Bimodality was discovered in the histogram of the distribution of positive Lyapunov exponents, calculated in the classical version, without renormalization of the shadow orbit, which makes it possible to implement a probabilistic method of GC classification. To construct the orbits of globular clusters, we used a gravitational potential model with a bar in the form of a triaxial ellipsoid, described in detail in the work of Bajkova et al., Izvestiya Glavnoi astronomicheskoi observatorii v Pulkove, {\bf 228}, 1 (2023). The following bar parameters were adopted: mass $10^{10} M_\odot $, semimajor axis length 5 kpc, bar viewing angle 25$^o$, rotation velocity 40 km s$^{-1}$ kpc$^{-1}$. To form the 6D-phase space required for orbit integration, the most accurate astrometric data to date from the Gaia satellite (EDR3) (Vasiliev, Baumgardt, 2021), as well as new refined average distances to globular clusters (Baumgardt, Vasiliev, 2021) were used. A classification is made of globular clusters with regular and chaotic dynamics. As the analysis showed, globular clusters with small pericentric distances and large eccentricities are most susceptible to the influence of the bar and demonstrate the greatest chaos. It is shown that the results of classifying globular clusters by the nature of their orbital dynamics, obtained using the various methods of analysis considered in the work, correlate well with each other.
}

\bigskip


\section*{INTRODUCTION}

This work is a continuation of a series of works by the authors of [1,2,3,4,5,6] devoted to the study of the orbital dynamics of globular clusters (GCs). Thus, the paper [1] presents a catalog of the orbits of 152 galactic globular clusters based on the latest astrometric data from the Gaia satellite (Gaia EDR3) [7], as well as new refined average distances [8]. In the work [2] an analysis was carried out (using the same data) of the influence of the galactic bar on the orbital motion of globular clusters in the central region of the Galaxy. For this task, 45 globular clusters were selected in the central galactic region with a radius of 3.5 kpc, 34 of which belong to the bulge and 11 to the disk. The list of selected GCs is given in Table~1. The orbits of globular clusters were obtained both in an axisymmetric potential and in a potential that included a bar model in the form of a triaxial ellipsoid. At the same time, the mass, rotation velocity and dimensions of the bar were varied. A comparison was made of such orbital parameters as apocentric and pericentric distances, eccentricity and maximum distance from the galactic plane.

The second stage of research aimed at studying the influence of the bar on the orbital motion of globular clusters was devoted to the problem of identifying objects captured by the bar using spectral dynamics methods [3,4,5,6].
The purpose of this work is to analyze the regularity/chaoticity of the orbits of all 45 selected GCs in the central region of the Galaxy using various methods, namely, methods for calculating the Highest Lyapunov Exponents (HLE) (in the classical version and in the version with renormalization of the $"$shadow$"$ orbit corresponding to the perturbed initial phase points, relative to the $"$reference$"$ orbit with given initial phase points), Poincar$\acute{e}$ sections, the frequency method based on the calculation of fundamental frequencies, as well as visual assessment from images of the reference and shadow orbits. All 45 GCs are listed in Table~1: the first column gives serial numbers, and the second column gives the GC names. As a bar model, we considered a model of an elongated triaxial ellipsoid with the most probable parameters known from the literature (for example, [9, 10]): mass $10^{10} M_\odot$, semi-major axis length 5 kpc, angle of inclination to the axis $X$ of 25$^o$, rotation velocity 40 km s$^{-1}$ kpc$^{-1}$.

It should be noted that we have already considered the problem of regularity/chaoticity of the orbital motion of a GC in the central region of the Galaxy in the work [5], but this consideration was very superficial and was based on only one method, namely the method of calculating the HLE in the classical version, i.e. without renormalization of the shadow orbit. The conclusions drawn in this work are not correct for all GCs in the sample. Therefore, this work sets the task of a more in-depth analysis using several of the most effective methods.

Since GCs in the central region of the Galaxy are subject to the greatest influence from the elongated rotating bar, the question of the nature of the orbital motion of the GC - regular or chaotic - is of great interest. For example, in the work [11] it is shown that the main share of chaotic orbits should be in the bar region.

In this work, we limit ourselves to considering the problem of identifying GCs with chaotic dynamics using the example of the gravitational potential, which we traditionally use to analyze the orbital motion of GCs [1,2,3,4,5].
The most detailed description of the Galactic gravitational potential model, which includes a three-component axisymmetric part (bulge, disk, halo) and an embedded central elongated bar, as well as astrometric data from the Gaia spacecraft necessary for the formation of a 6D phase space for orbit integration, is given in already mentioned work [2]. A selection of 45 globular clusters is also carried out there. The work [2] also provides all the necessary literary references. Therefore, in this paper we omit all technical details associated with the description of the gravitational potential, both axisymmetric and eaxisymmetric, data, as well as the selection of globular clusters. Here we consider only methodological issues related to the analysis of the regularity/chaoticity of the orbital motion of a GC, based on the methods listed above, to which Section 1 is devoted. In Section 2, the obtained results of GC classification are compared. In the CONCLUSIONS section the main results of the work are formulated.

\section{Methods and results of analysis of regularity/chaoticity of GC orbital dynamics}

\subsection{Direct method for calculating HLE - probabilistic method}

The values of the Highest Lyapunov Exponents (HLE) are calculated by the $"$shadow$"$ trajectory method using the following formula ([12]):

\begin{equation}
\label{Lyap}
L(n)=\frac{1}{n \delta t} \sum_{i=1}^{n} \ln{\frac{D_i}{D_{i-1}}},
\end{equation}
where $D_i$ is the distance in three-dimensional space between the reference and shadow phase points at the $i$-th integration step, $D_0$ is the length of the displacement vector at the initial time, $\delta t$ is a fixed value of the integration step along time. The true value of the HLE is equal to the limit $L(n)$ at $n \rightarrow \infty$, i.e. as the integration time tends to infinity, and $D_0 \rightarrow 0$. In practice, $L(n)$ obtained for a large value of $n$ is taken as the HLE value. In this case, non-zero positive values of the HLE indicate chaotic, and zero and negative values indicate a regular nature of the movement.

In the classic version, i.e. in the absence of renormalization of the shadow orbit, the formula (\ref{Lyap}) is transformed into a simpler one:

\begin{equation}
\label{Lyapp_1}
L(n)=\frac{1}{n \delta t} \ln{\frac{D_n}{D_0}}.
\end{equation}

\begin{figure*}
{\begin{center}

              \includegraphics[width=0.3\textwidth,angle=-90]{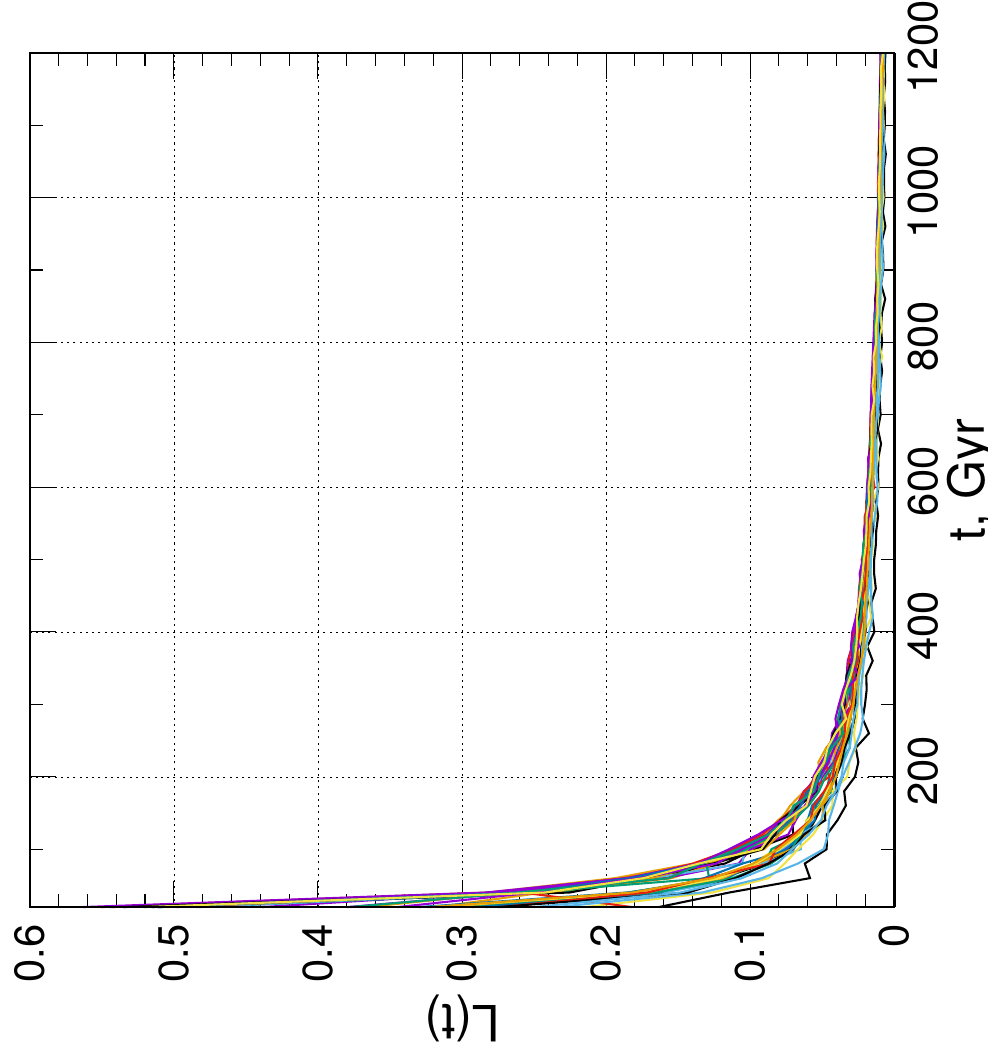}
              \includegraphics[width=0.3\textwidth,angle=-90]{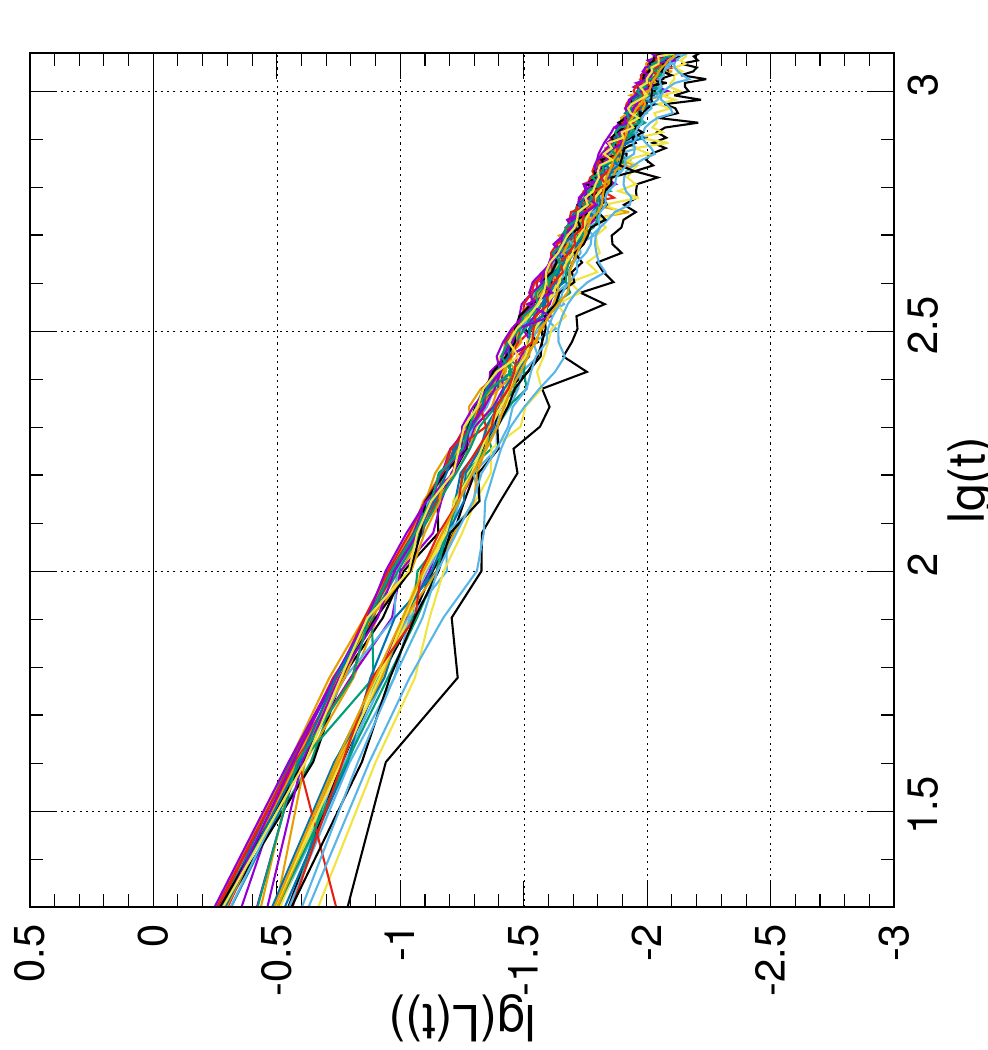}\

              \bigskip

               \includegraphics[width=0.3\textwidth,angle=-90]{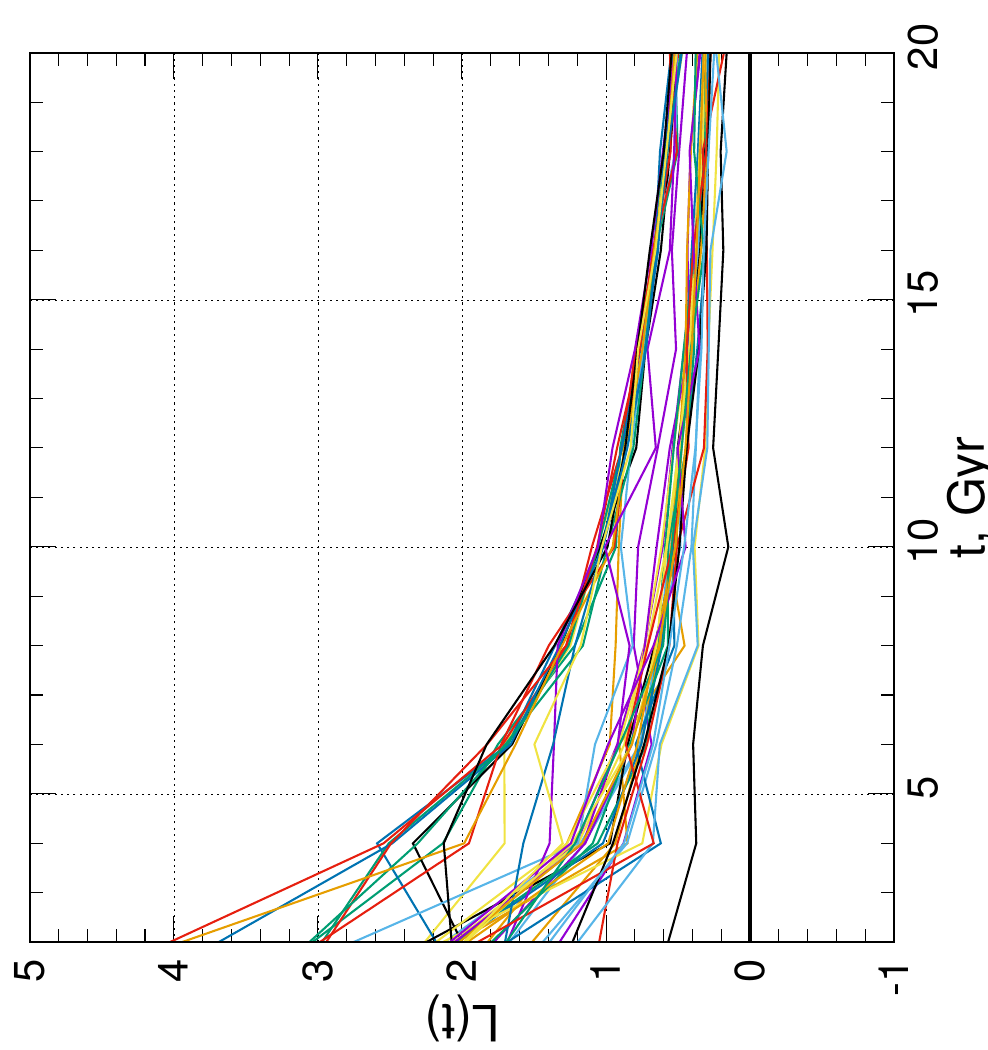}
               \includegraphics[width=0.3\textwidth,angle=-90]{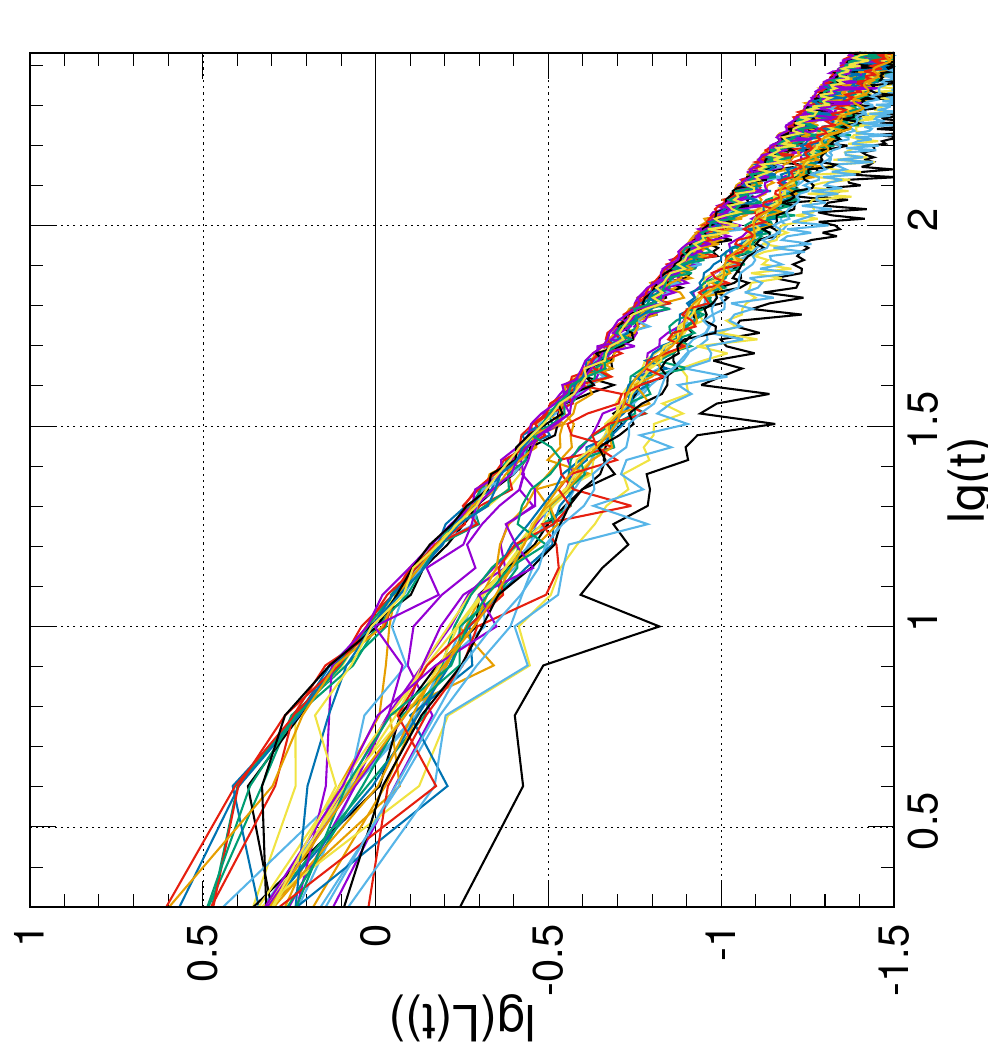}\

 \bigskip

\caption{\small Calculation of HLE approximations without renormalization of the shadow orbit as a functions of time for 45 GCs. The top row is the maximum time interval of 1200 billion years, the bottom row is 20 billion years; on the left panels the functions are presented on a linear scale, on the right - on a logarithmic scale.}
\label{Lyap1}
\end{center}}
\end{figure*}

Dependences of $L(n)$ obtained for the orbits of all 45 globular clusters with the following perturbation of the phase initial point: $x_1=x_0+x_0\times 0.00001,~y_1=y_0+y_0\times 0.00001,~z_1=z_0+z_0\times 0.00001$, are shown in Fig.~\ref{Lyap1}. In the top row, functions $L(n)$ are presented over a time interval of 1200 billion years, in the bottom row - 20 billion years; on the left panels the functions $L(n)$ are given on a linear scale, on the right panels - on a logarithmic scale. It is clear from the graphs that the approximations of the Lyapunov exponents are positive and tend to zero as $n$ increases. In addition, the set of 45 functions $L(n)$ is divided into two families, this is especially clear for the graphs on a shorter time interval of 20 billion years. This indicates bimodality in the distribution of HLE approximations.

\begin{figure*}
{\begin{center}

               \includegraphics[width=0.3\textwidth,angle=-90]{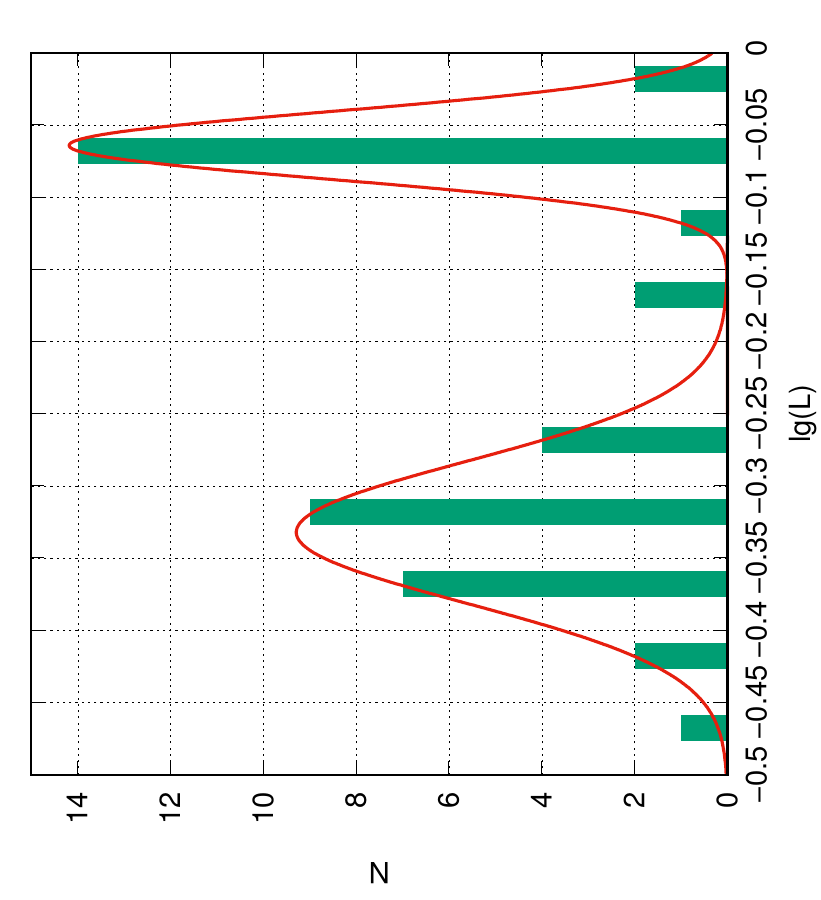}
               \includegraphics[width=0.3\textwidth,angle=-90]{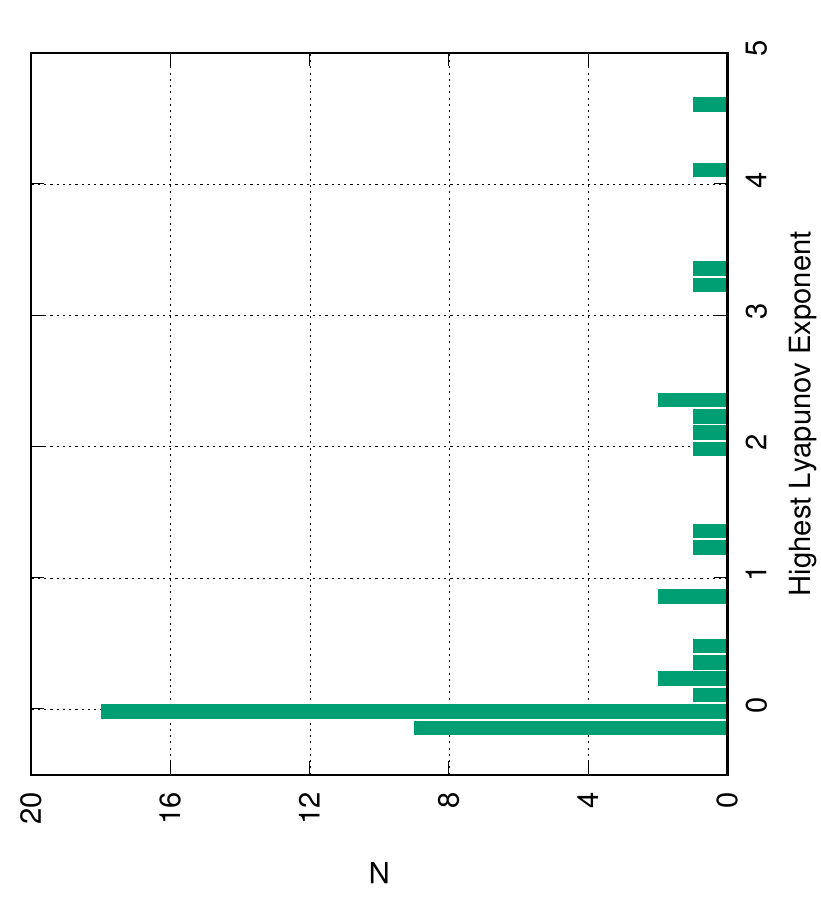}\

\bigskip

\caption{\small Left: histogram of the distribution of HLE approximations without renormalization of the shadow orbit for 45 GCs at $t=12$ billion years. Approximation of the histogram by two Gaussians (red line) makes it possible to implement a probabilistic method for separating GCs with regular (left Gaussian) and chaotic orbits (right Gaussian). Right: histogram of the distribution of HLE approximations obtained over an interval of 120 billion years with renormalization of the shadow orbit.}
\label{Lyap2}
\end{center}}
\end{figure*}

The histogram of the distribution of HLE approximations for $t=12$ billion years is shown in Fig.~\ref{Lyap2}, where this bimodality is clearly visible. By approximating the histogram with two Gaussian distributions using the least squares method and calculating for each GC the probability of belonging to one or another distribution, we obtain a probabilistic method for dividing the entire set of GCs into two subsets with relatively small (left Gaussian) and large HLE values (right Gaussian), indicating different degrees discrepancies between the reference and shadow orbits. As a result of applying the probabilistic method, we received two lists: the first list of 26 GCs with minimal differences between the reference and shadow orbits (NGC 6266, Terzan 4, Liller 1, NGC 6380, Terzan 1, Terzan 5, Terzan 6, Terzan 9, NGC 6522 , NGC 6528, NGC 6624, NGC 6637, NGC 6717, NGC 6723, Terzan 3, NGC 6304, Pismis 26, NGC 6569, E456-78, NGC 6540, Djorg 2, NGC 6171, NGC 6316, NGC 6388, NGC 653 9, NGC 6553) and a second list of 19 GCs with noticeable discrepancies between the reference and shadow orbits at long time intervals (NGC 6144, E452-11, NGC 6273, NGC 6293, NGC 6342, NGC 6355, Terzan 2, BH 229, NGC 6401, Pal 6, NGC 6440, NGC 6453, NGC 6558, NGC 6626, NGC 6638, NGC 6642, NGC 6256, NGC 6325, NGC 6652). We assume that the first list includes GCs with regular orbits, and the second - those with chaotic ones.

\begin{figure*}
{\begin{center}

               \includegraphics[width=0.3\textwidth,angle=-90]{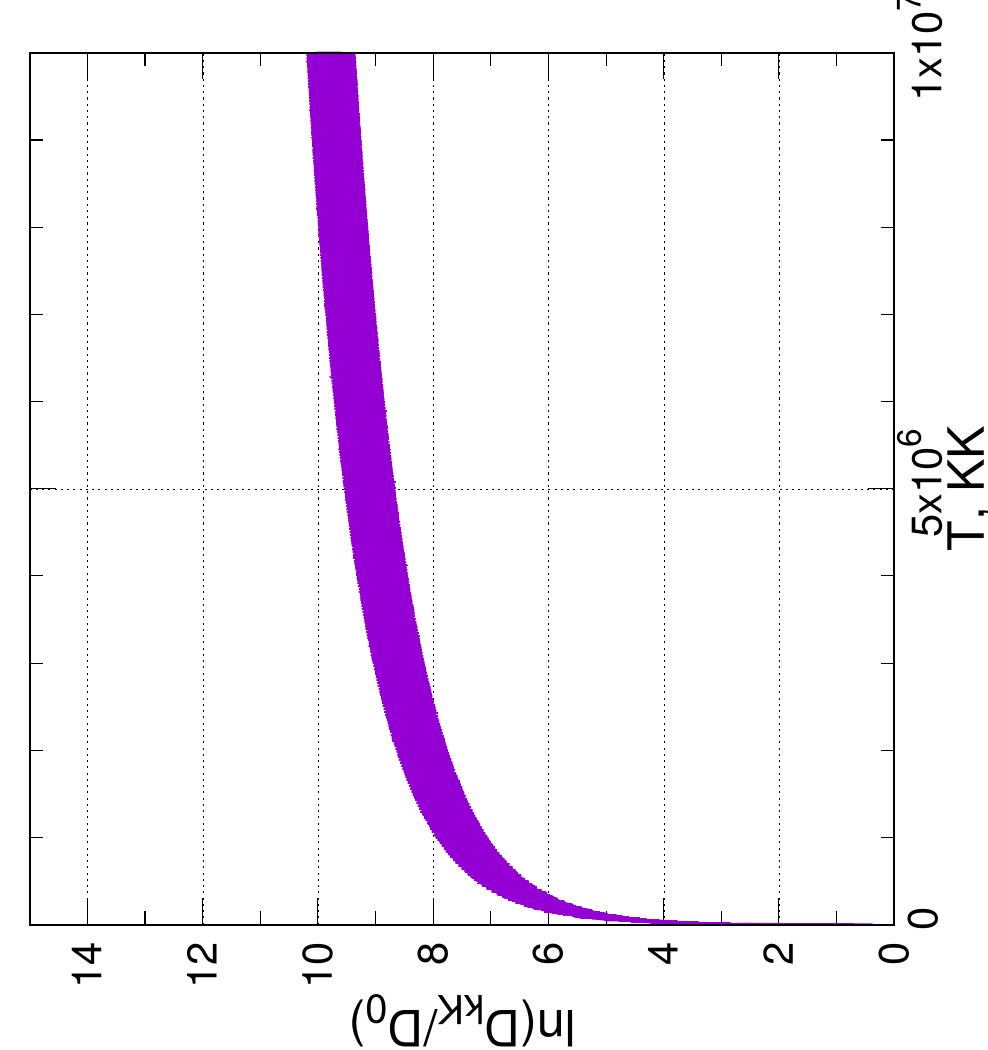}
               \includegraphics[width=0.3\textwidth,angle=-90]{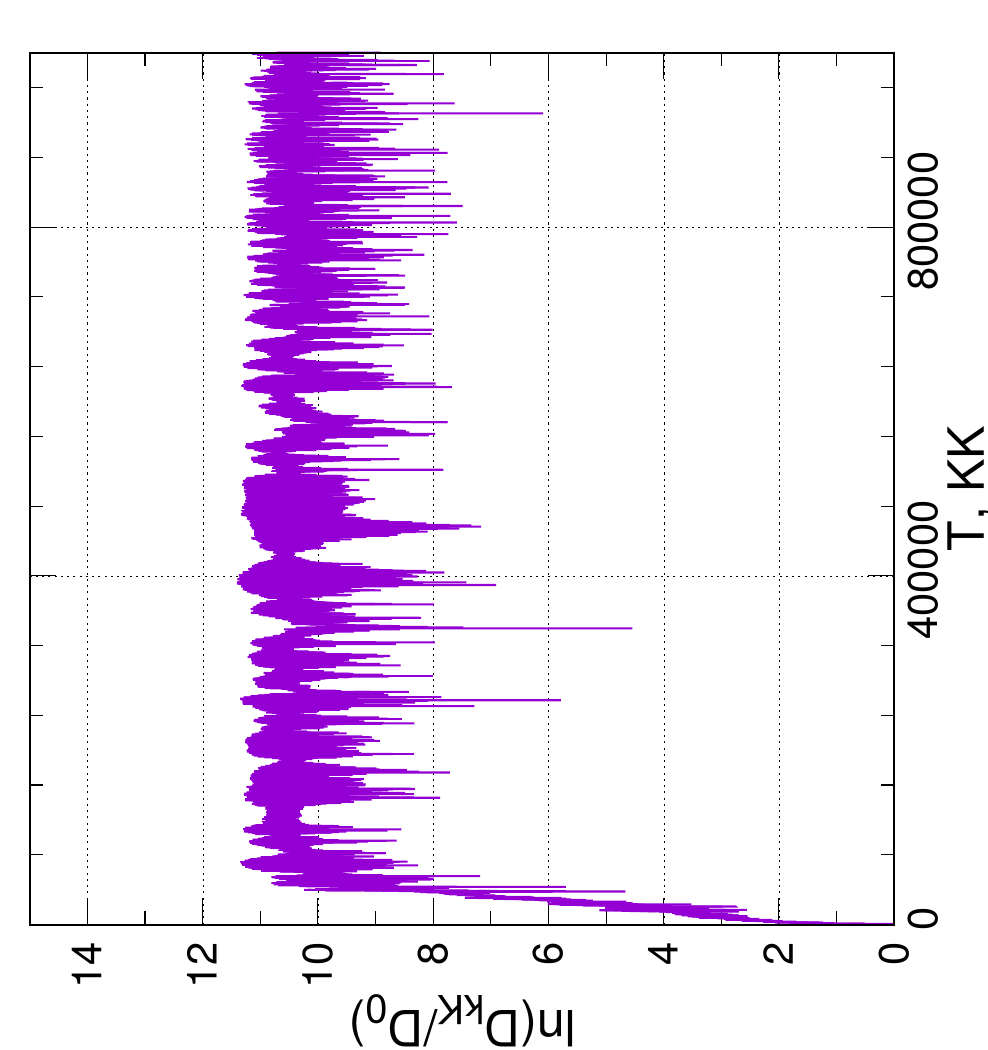}\

               \includegraphics[width=0.3\textwidth,angle=-90]{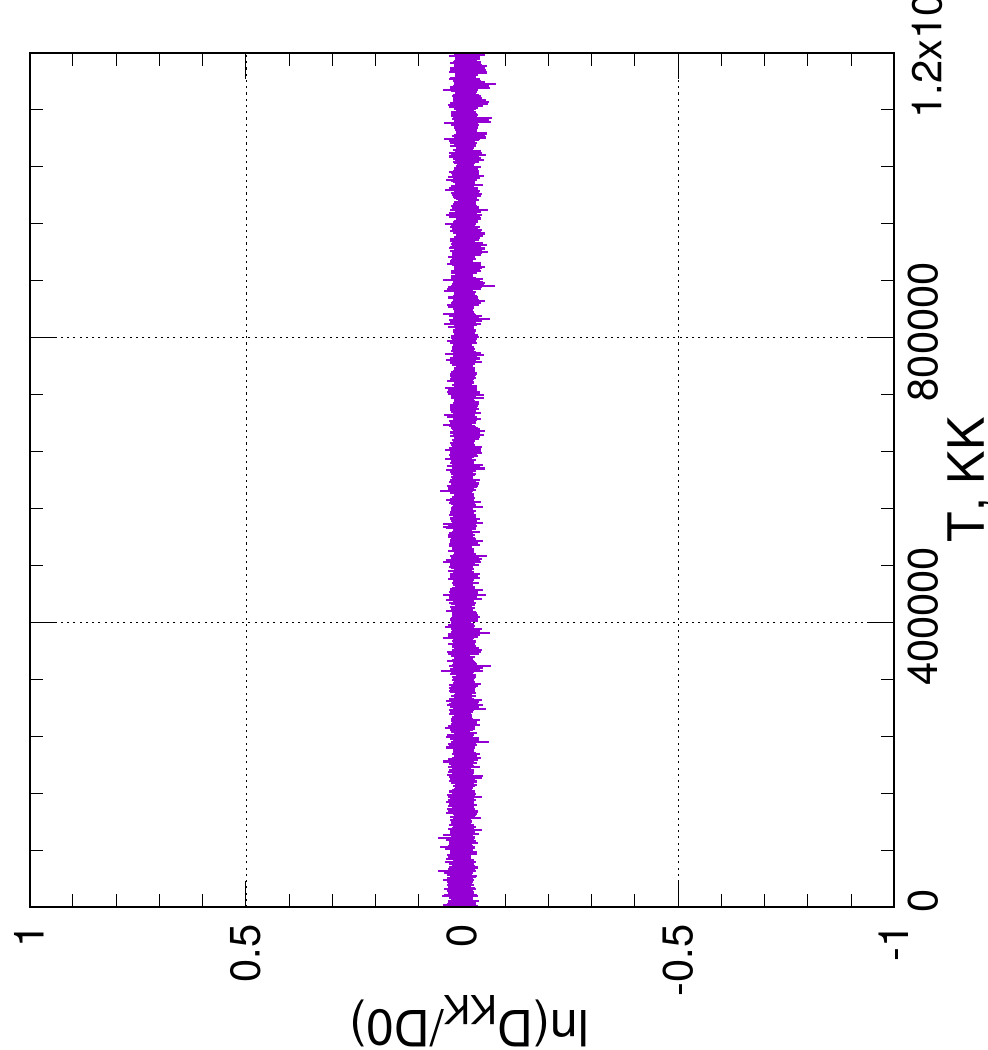}
               \includegraphics[width=0.3\textwidth,angle=-90]{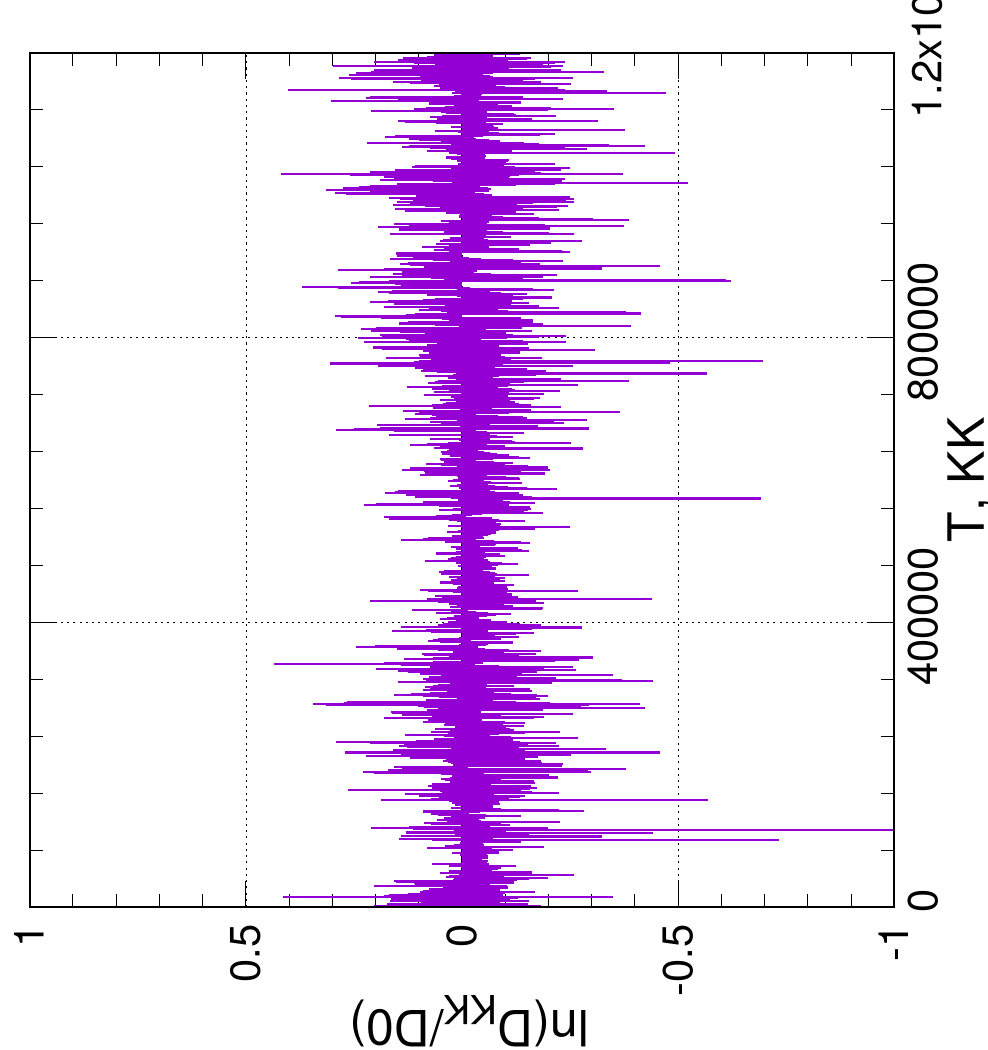}\

\bigskip

\caption{\small  Typical examples of the $\ln (\frac{D_t}{D_0})$ dependence for supposedly regular (upper left panel) and chaotic orbits (upper right panel) in the case of calculating the HLE without renormalizing the shadow orbit. Here, the time $t$ is represented by discrete samples $KK$, the time step size is 0.1 million years, i.e. $t=0.1\times KK$ million years. Examples are given for NGC 6266 (regular orbit) and NGC 6355 (chaotic orbit). The $\ln (\frac{D_t}{D_0})$ dependences in the case of renormalization of the shadow orbit are shown in the lower left panel for NGC 6266 and in the lower right panel for NGC 6355. The upper row of dependences is shown for a time interval of 100 billion years, the lower row -- 120 billion years. }
\label{Lyap3}
\end{center}}
\end{figure*}

In order to explain the resulting bimodality of the HLE approximations, we plotted the dependences of $f(n)=\ln{\frac{D_n}{D_0}}$ on $n$ for each GC and obtained two types of graphs typical for GCs from the first and second lists , shown in the left and right upper panels of Fig.~\ref{Lyap3}, respectively. We see the different nature of the obtained dependences $f(n)$. In the first case, the curve $f(n)=\ln{\frac{D_n}{D_0}}$ has a smoother (with minimal spread) and flatter shape than in the second case, while as $n$ tends to infinity, the curve continues to grow, albeit with a strong slowdown. In the second case, the values of the function $f(n)=\ln{\frac{D_n}{D_0}}$ already at relatively small values of $n$ quite sharply reach large values, albeit with a large scatter, and with a further increase in $n$ reach saturation. At the same time, the scope of the curve in the first case over the considered time interval of 1200 billion years is slightly lower than in the second case.
A graphic illustration for all 45 GCs is given in Fig.~10 in the sixth horizontal row of panels from the top. Now the nature of the $"$hyperbolic$"$ dependence $L(n)$ (Fig.~\ref{Lyap1}), calculated by the formula (\ref{Lyap}) and asymptotically tending to zero as $n$ increases, is also clear only for supposedly regular, but also chaotic orbits, and the reason for the bimodality of the distributions of the HLE GC approximations of our sample is clear.

Note that here we have considered the case of calculating HLE approximations without renormalizing the shadow orbit (formula (\ref{Lyapp_1})). Naturally, this algorithm, due to the obtained dependencies $f(n)=\ln{\frac{D_n}{D_0}}$, when a very large deviation of the reference and shadow phase points is observed, cannot serve for the correct calculation of the HLE. When calculating using the formula (\ref{Lyap}), the position of the shadow phase point must be periodically renormalized
relative to the reference one by the distance $D$ between them, so that this distance is always
relatively small. However, the algorithm without renormalization is quite suitable for separating regular and chaotic orbits due to the difference in the dependencies $f(n)=\ln{\frac{D_n}{D_0}}$. In what follows, we will simply call this algorithm probabilistic, since it divides the GC according to the principle of maximum probability of belonging to a set of GCs with regular or chaotic motion.

The designations for the classification of GCs with regular (Regular) (R) and chaotic (Chaotic) (C) motion, obtained by the probabilistic method, are listed in the third column of Table~1.

\subsection{Calculation of HLE with shadow orbit renormalization}

To correctly calculate the HLE, the displacement vector of the shadow phase point from the reference point was renormalized at each small time interval $\Delta t$ so that the phase point of the shadow orbit shifted along the vector separating the phase points of the reference and shadow orbits back to the original value of the displacement modulus $D_0$ (see, for example, Fig.9.9 in [13]). If $n_t$ steps are taken, then the HLE estimate is given by the modified formula ([13,14])
\begin{equation}
\label{Lyap_2}
L(t)=\frac{1}{n_t \Delta t} \sum_{i=1}^{n_t} \ln{\frac{D_i}{D_0}},
\end{equation}
where $t=n_t\Delta t$ is the total integration interval.

For the GCs of our sample, we experimentally established that the optimal size of the renormalization interval $\Delta t$ is a value equal to $(30-50)\delta t$, where $\delta t$ is the time integration step of the orbit (see. also the formula (\ref{Lyap})). In our case, $\delta t=0.0001$ billion years for all GCs. The value of the renormalization interval $\Delta t$ for each GC was selected individually from the given interval. We calculated the distances between the reference and shadow orbits as a function of time in units of $\ln (\frac{D_t}{D_0})$ at each discrete point $t_n=n\times\delta_t$ over a time interval of 120 billion years in the case of renormalization of the shadow the orbits shown in the bottom row of panels in Fig.~\ref{Lyap3} for NGC 6266 on the left and for NGC 6355 on the right. Comparison with similar dependences obtained without renormalization of the shadow orbit (top row of panels in Fig.~\ref{Lyap3}) shows that renormalization of the shadow orbit led to a significant decrease in the distance between the phase points of the reference and shadow orbits. This made it possible to correctly calculate the HLE approximations. The unit of measurement for HLE in our case is a value inversely proportional to the time of 1 billion years, i.e. 1/Gyr. The values of the HLE approximations, calculated using the formula (~\ref{Lyap_2})) with renormalization of the shadow orbit relative to the reference one in the $\Delta_t$ interval over a total integration time of 120 billion years, amounted to -0.017/Gyr $(<0)$ for NGC 6266 and 2.257/Gyr $(>0)$ for NGC 6355, which indicates a regular orbit in the first case and chaos in the second one.

We calculated the HLE approximations with renormalization of the shadow orbit for all 45 GCs in our sample. The corresponding histogram of the distribution of HLE values is shown in the right panel of Fig.~\ref{Lyap2}. The list of GCs with regular orbits includes objects with approximation values HLE$<0$, and the list of GCs with chaotic orbits includes GCs with HLE values$>0$. The values of the HLE approximations along with the designations of the orbits - regular (R) or chaotic (C) are given in the fifth column of Table~1. Note that the present classification is somewhat different from the previous one, obtained by the probabilistic method (see also the third column of Table~1). The correlation coefficient between the results of orbit classification by these two methods is $K_c=0.60$ (Table~3). A detailed comparison of the results of analyzing the regularity of GC orbits obtained by various methods will be given below in Section 2.

\subsection{MEGNO}
A description of MEGNO (Mean Exponential Growth factor of Nearby Orbits) can be found in the works [12,14]. MEGNO is one of the most widely used methods for identifying chaos in various problems of celestial mechanics, and at significantly shorter times than HLE.
When analyzing the regularity of orbits, we use the MEGNO property that in the case of a regular trajectory $M(t)\rightarrow 2$ for $t \rightarrow \infty$.

\begin{figure*}
{\begin{center}

              \includegraphics[width=0.3\textwidth,angle=-90]{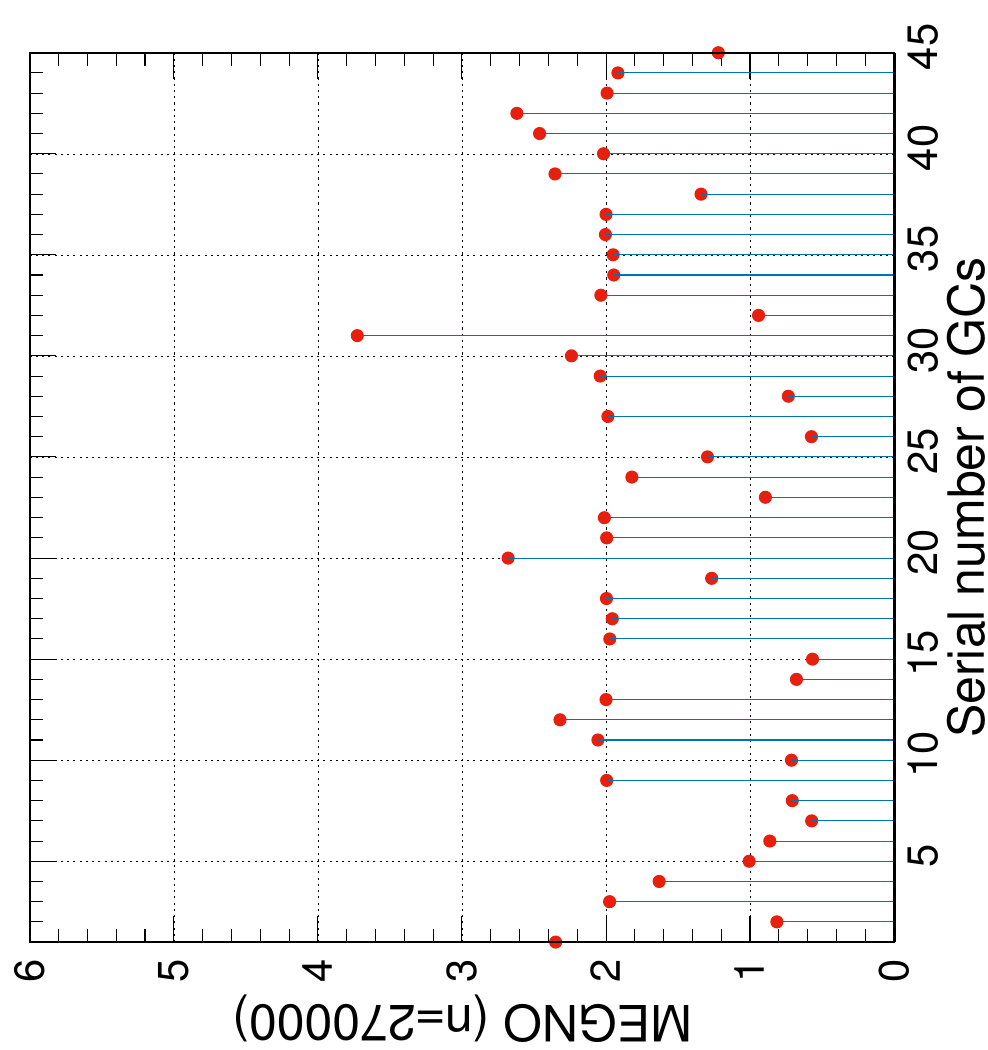}
              \includegraphics[width=0.3\textwidth,angle=-90]{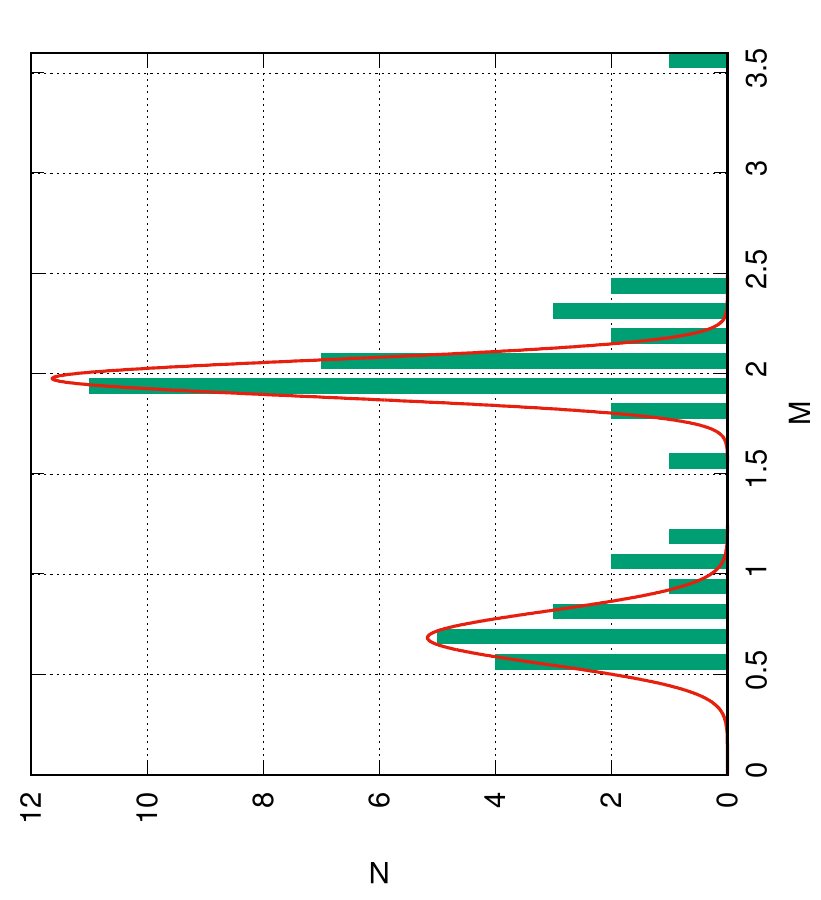}
              \includegraphics[width=0.3\textwidth,angle=-90]{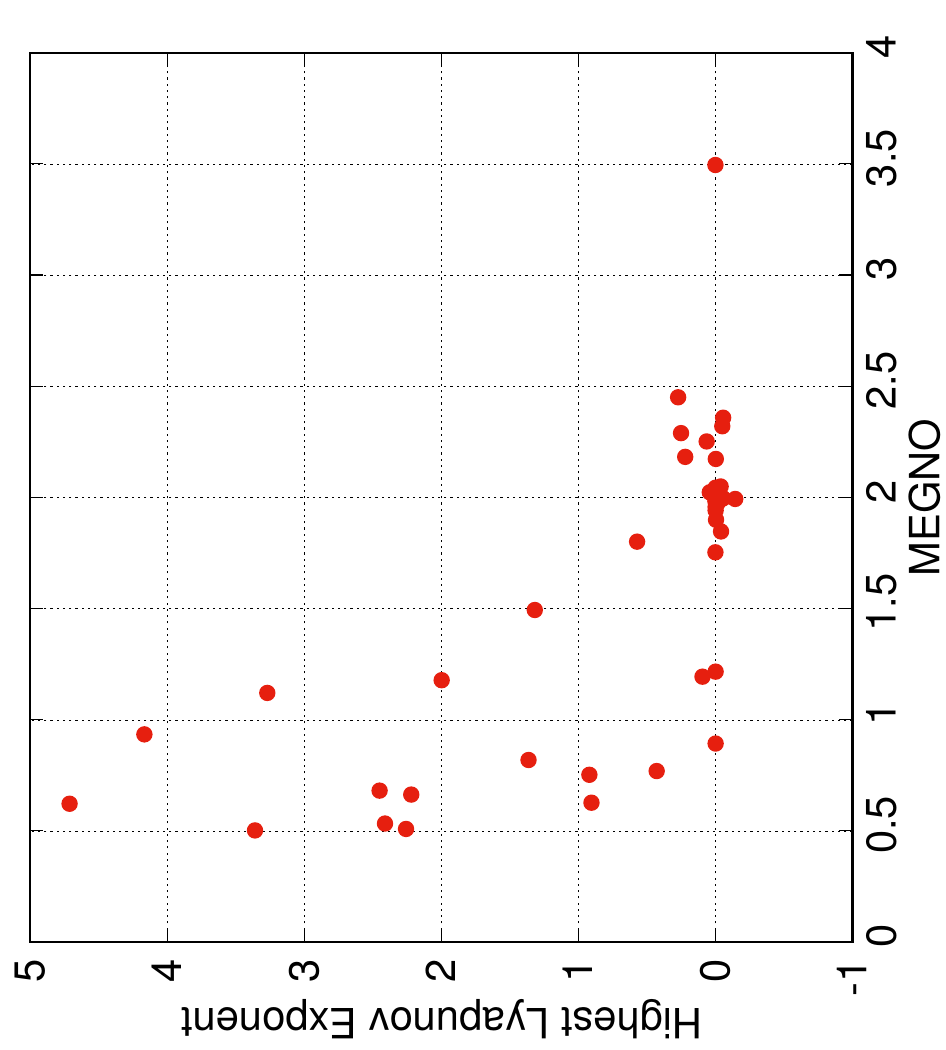}\

\bigskip

\caption{\small MEGNO approximations for 45 GCs on a time interval of 270 Gyr (left panel). Histogram of the distribution of MEGNO approximations and its approximation by two Gaussians (red line), which makes it possible to implement a probabilistic method for separating GCs with regular and chaotic orbits (middle panel). Diagram $"$MEGNO - HLE$"$ (right panel).}
\label{Lyap4}
\end{center}}
\end{figure*}

In this work, we use a method well described in [12] and proposed by Breiter et al. in [15] for the case of numerical integration with a constant step, which is as follows: the value of the parameter MEGNO at the n-th step of integration determined by the formula:
\begin{equation}
\label{megno}
m(n)=\frac{n-1}{n} m(n-1) + 2 \ln{\frac{D_n}{D_{n-1}}}.
\end{equation}
For the time-averaged value MEGNO we have:
\begin{equation}
\label{megno1}
M(n)=\frac{1}{n} ((n-1)M(n-1) + m(n)),
\end{equation}
assuming $m(0)=0$ and $M(0)=0$.

To correctly calculate MEGNO values in the same way as for correct
calculation of HLE values it is required renormalization of the position of the shadow phase point relative to the reference point by the distance $D$ between them. But since estimates of the magnitude of the HLE using MEGNO are characterized by less reliability than the direct calculation of the HLE (see paper [12] and references there), we limited ourselves to calculating MEGNO without renormalization of the shadow orbit in order to obtain only a qualitative result suitable for separating regular and chaotic orbits similar to the probabilistic method based on calculating the HLE without renormalization (see Section 1.1).

The result of applying the formulas (\ref{megno})-(\ref{megno1}) to our sample over a time interval of 270 Gyr (n=270000, $\delta t=0.001$ Gyr) is shown in Fig.~\ref{Lyap4}. The values of the MEGNO appoximations for 45 GCs are shown in the left panel (the abscissa axis shows the serial numbers of the GCs in accordance with Table~1). The histogram of the distribution of MEGNO approximations and its approximation by two Gaussians (red line), which makes it possible to implement a probabilistic method for separating GCs with regular (right Gaussian centered at the point $M\approx2$) and chaotic orbits (left Gaussian), are shown in the middle panel.

The lists of GCs with regular and chaotic orbits, obtained on the basis of MEGNO and HLE without normalization of the shadow orbit, practically coincide (with the exception of NGC 6144). See also the fourth column of Table~1, where the values of the MEGNO approximations for each GC and the orbit classification designations are indicated as (R) or (C). The correlation coefficient between these classification results is $K_c=$0.95 (see Table~3).

The $"$MEGNO - HLE$"$ diagram with renormalization of the shadow orbit (in the HLE method) is shown in the right panel of Fig.~\ref{Lyap4}. The correlation coefficient between the calculated values of the MEGNO and HLE approximations is $K_c=$0.70 (see Table~2).

\subsection{Poincar$\acute{e}$ sections}

One of the methods for determining the nature of motion (regular or chaotic) is the analysis of Poincar$\acute{e}$ sections [13]. The algorithm we used to construct the mappings is as follows:

1. We consider the phase space $(X,Y,V_x,V_y)$.

2. We eliminate $V_ Y$ using the Jacobi integral conservation law and go to the space $(X,Y,V_x)$.

3. We define the plane $Y=0$; we denote the points of intersection with the orbit on the plane $(X,V_x)$. We take only those points at which $V_y>0$.

The phase space $(Y,Z,V_y,V_z)$ or $(R,Z,V_R,V_z)$ can be considered similarly. Then the Poincar$\acute{e}$ sections will be reflected on the plane $(Y,V_y)$ or $(R,V_R)$, respectively.

If the intersection points of the plane add up to a continuous smooth line (or several separated lines), then the movement is considered regular. In the case of chaotic motion, instead of being located on a smooth curve, the points fill a two-dimensional region of phase space, and sometimes the effect of points sticking to the boundaries of islands corresponding to ordered motion [13] occurs.

 It should be noted that for non-axisymmetric potential models, which includes the potential considered in this work, which includes a rotating central bar, the Poincar$\acute{e}$ sections have a more complex structure than in the case of an axisymmetric model. If in the case of axisymmetric models for regular orbits the Poincar$\acute{e}$ sections usually represent a straight line, then in the case of non-axisymmetric models for many orbits more complex patterns are obtained. It would be incorrect to call such orbits chaotic, since obvious patterns are observed in the arrangement of points, but they may no longer form a single line. Thus, the task of dividing orbits into regular and chaotic on the basis of Poincar$\acute{e}$ sections becomes noticeably more complicated and is not without subjectivity. Therefore, it is of great importance to involve, along with Poincar$\acute{e}$ sections, other methods of analysis and make a decision about the nature of the movement of objects based on the results of using several independent approaches.

\begin{figure*}
{\begin{center}
                \includegraphics[width=0.3\textwidth,angle=-90]{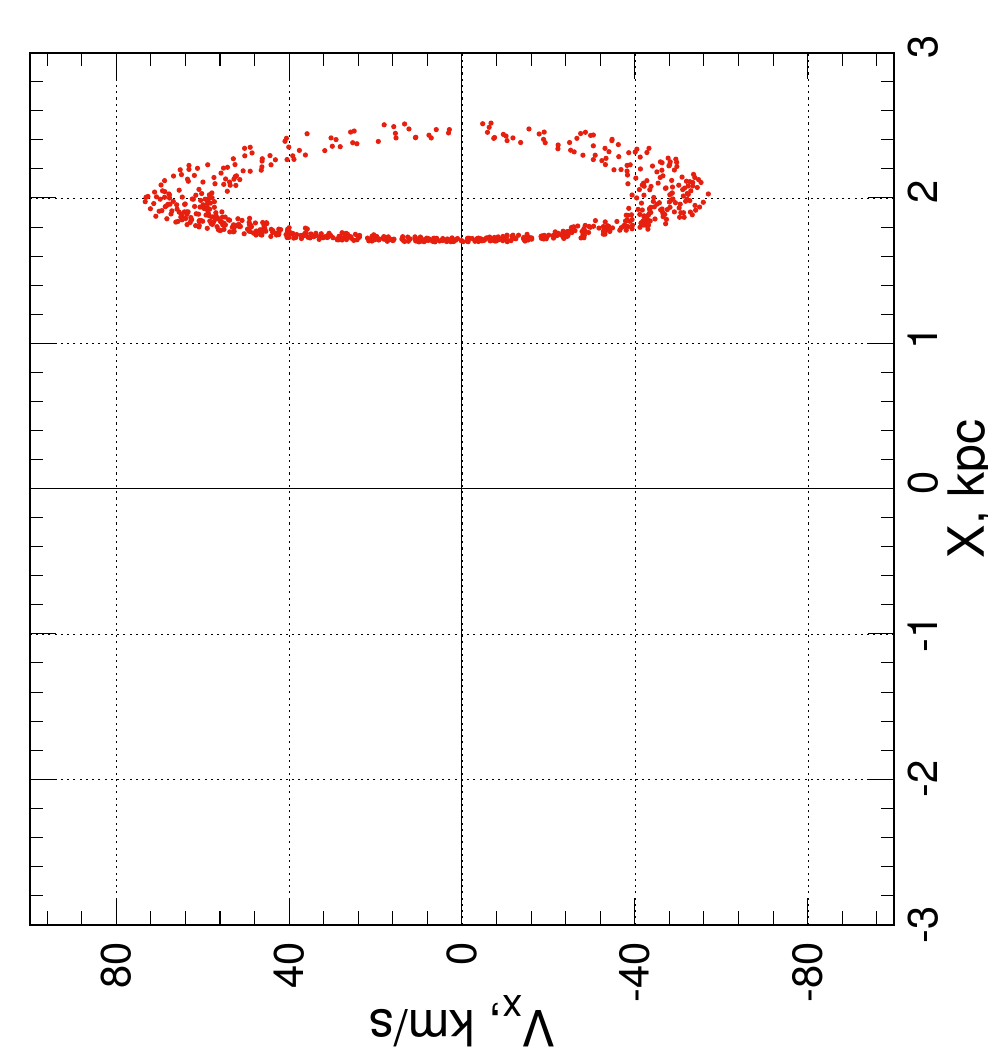}
                \includegraphics[width=0.3\textwidth,angle=-90]{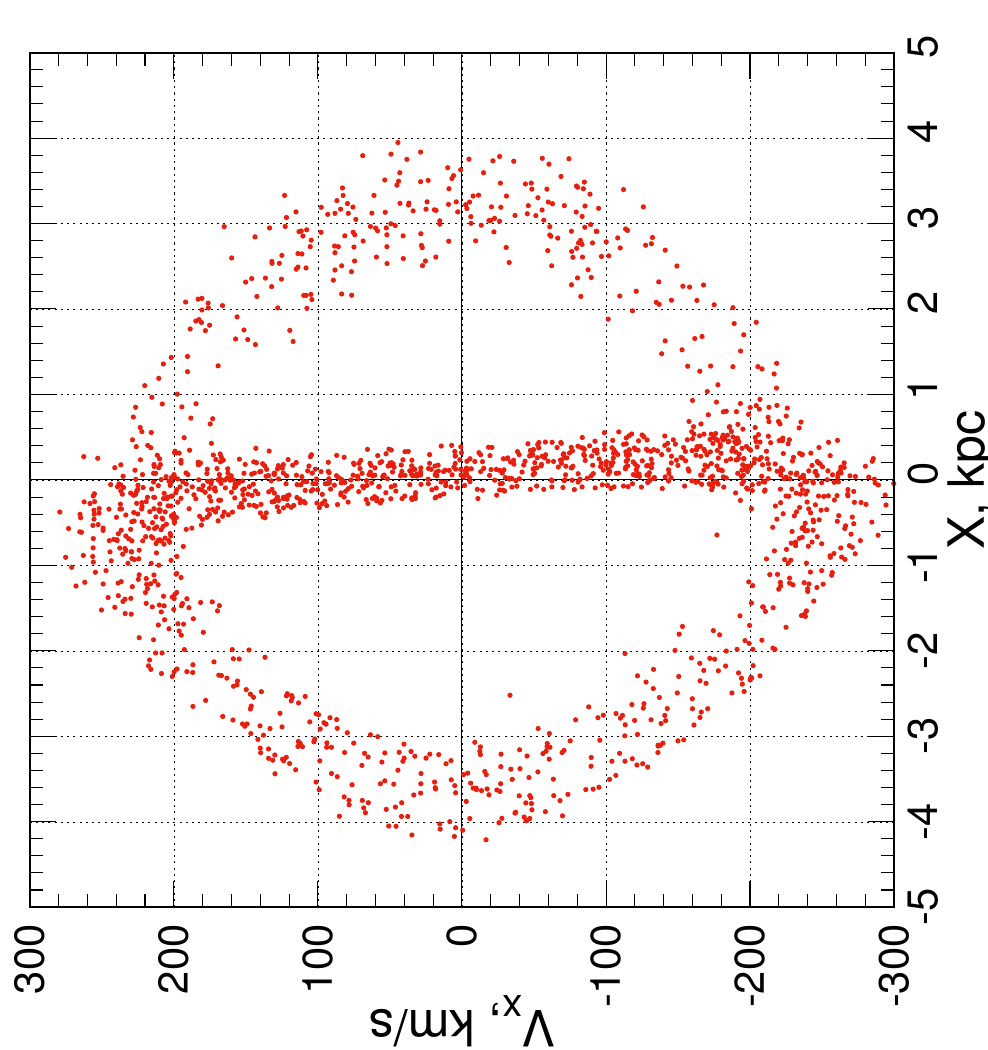}\

\bigskip

\caption{\small Poincar$\acute{e}$ sections. Left: for NGC 6266 (regular orbit), right: NGC 6355 (chaotic orbit).}
\label{Lyap5}
\end{center}}
\end{figure*}

Fig.~\ref{Lyap5} shows an example of Poincar$\acute{e}$ sections for regular (NGC 6266) and chaotic (NGC 6355) motion.

 We calculated the Poincar$\acute{e}$ sections $(X,V_x)$ for all 45 GCs and, by visual analysis of them, tried to determine the nature of the GC motion as (R) or (C) with the greatest possible objectivity, sometimes (in complex cases) using the results of more an objective (in our opinion) method based on frequency analysis, which is discussed in the next section. The classification results are reflected in the sixth column of table~1, and a graphical representation of the sections is shown in Fig.~10 in the fifth (from top) horizontal row of panels.

\subsection{Frequency method}

Another way to study orbital regularity/chaoticity involves the use of orbital frequencies [16,17] (see section 3.1 in [17]) have shown that it is possible to measure orbital stochasticity based on the shift of fundamental frequencies determined over two consecutive time intervals. For each frequency component $f_i$, a parameter called frequency drift is calculated:
\begin{equation}
\label{freq}
\lg(\Delta f_i)=\lg|\frac{\Omega_i(t_1)-\Omega_i(t_2)}{\Omega_i(t_1)}|,
\end{equation}
where $i$ defines the frequency component in Cartesian coordinates
(i.e. $\lg(\Delta f_x), \lg(\Delta f_y)$ and $\lg(\Delta f_z)$). The largest value of these three frequency drift parameters $\lg(\Delta f_x)$ is then assigned to the frequency drift parameter $\lg(\Delta f)$. The higher the $\lg(\Delta f)$ value, the more chaotic the orbit. However, as shown in [17], the accuracy of frequency analysis requires at least 20 oscillation periods to avoid classification errors.

\begin{figure*}
{\begin{center}
               \includegraphics[width=0.3\textwidth,angle=-90]{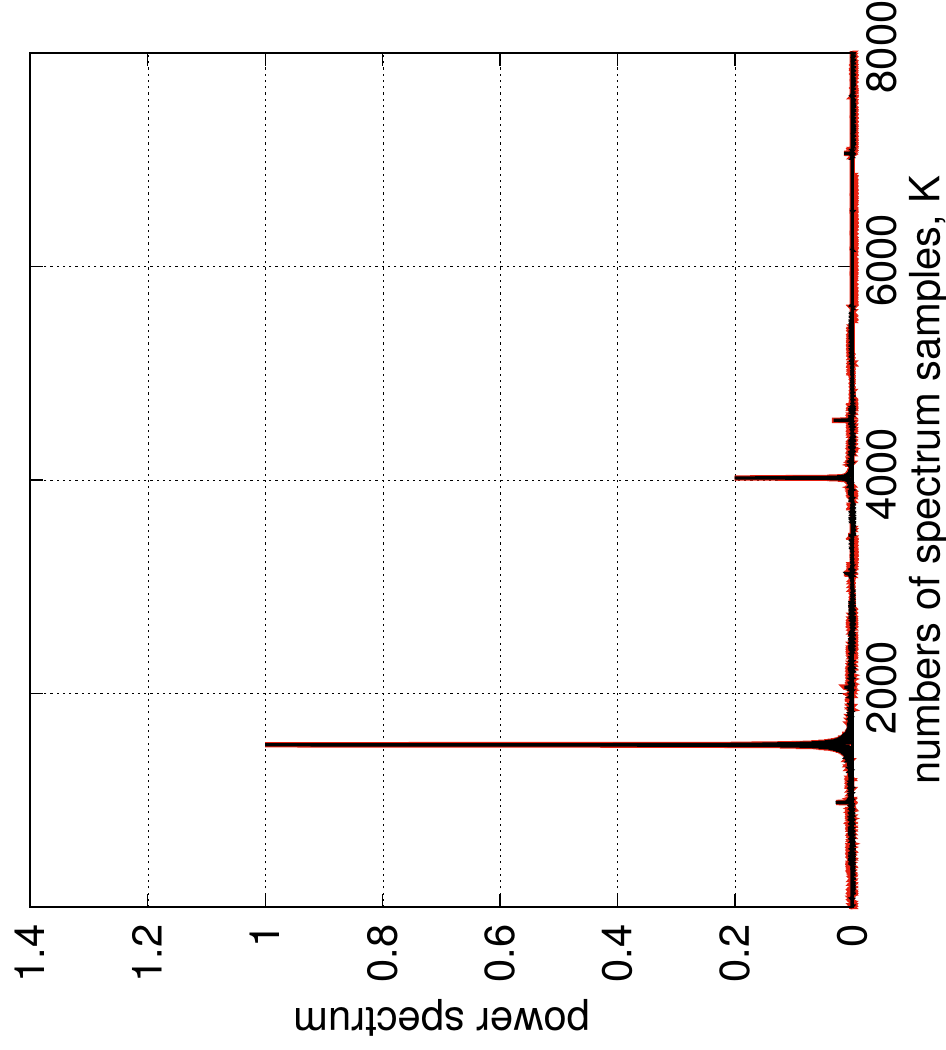}
               \includegraphics[width=0.3\textwidth,angle=-90]{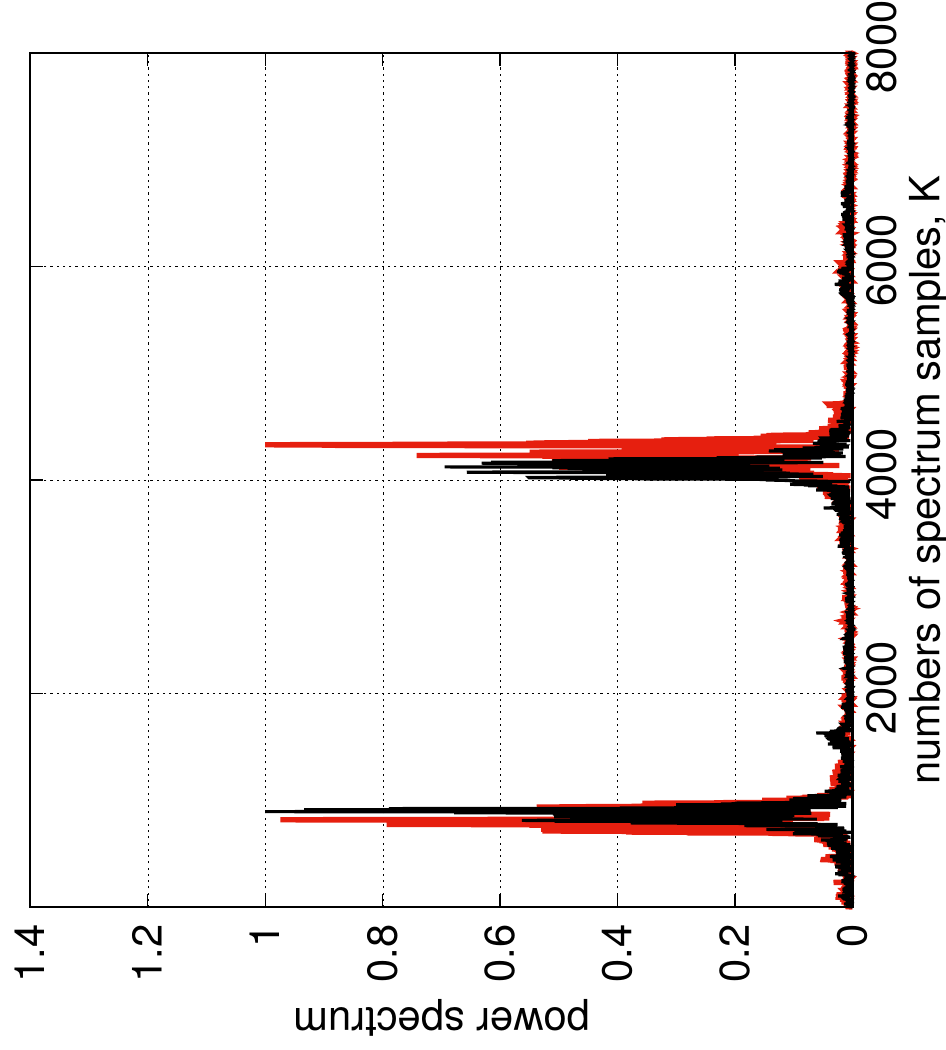}\

               \includegraphics[width=0.3\textwidth,angle=-90]{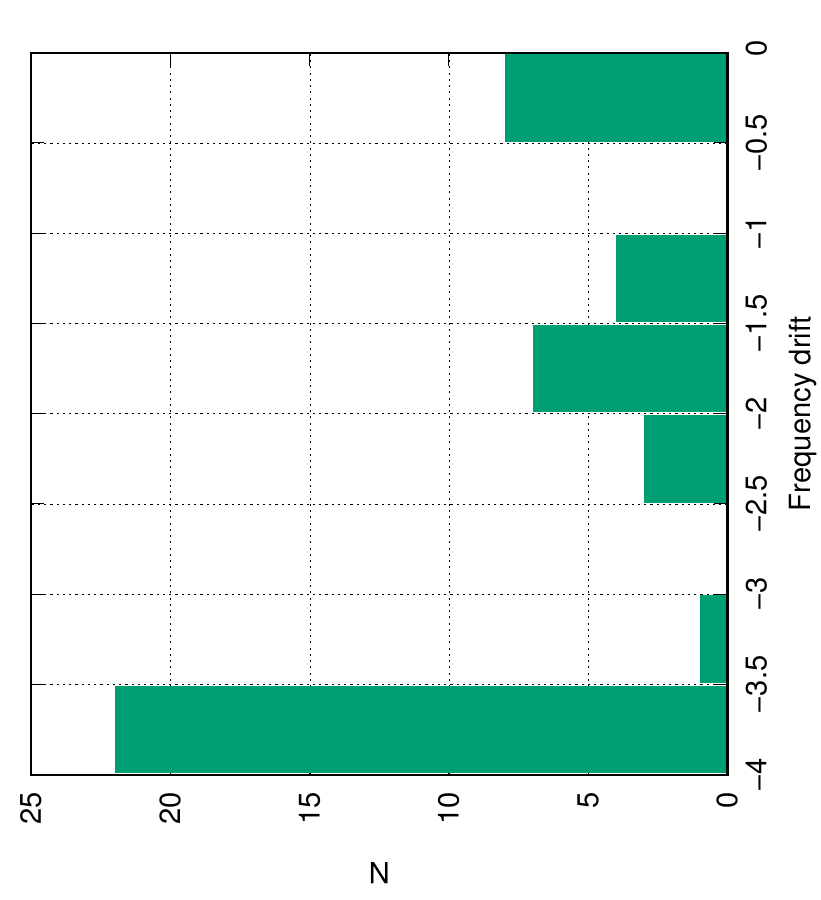}
               \includegraphics[width=0.3\textwidth,angle=-90]{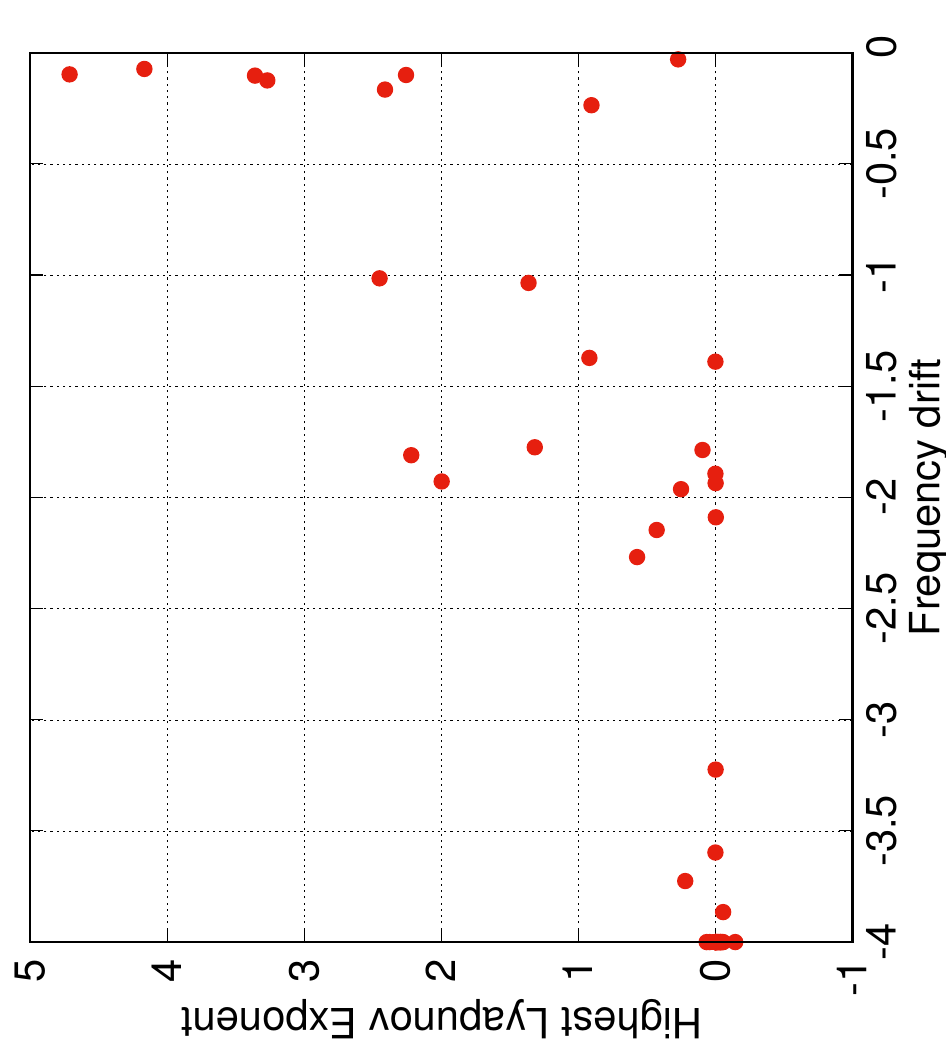}\

\bigskip

\caption{\small Top panels: illustration of the frequency method, the amplitude spectrum of the first half of the time sequence is shown in red, and the second half in black. The left panel refers to NGC 6266 with a regular orbit, the right panel refers to NGC 6355 with a chaotic orbit.
Bottom panels: histogram of frequency drift parameter distribution
$\lg(\Delta f)$ (left panel), $"$frequency drift -- HLE$"$ diagram (right panel).}
\label{Lyap6}
\end{center}}
\end{figure*}

We calculated the frequency drift parameter $\lg(\Delta f)$ for all 45 GCs, from which we determined the nature of their movement as (R) or (C). The series $x(t_n), y(t_n), z(t_n)$ were determined over the time interval [0, 120] billion years. The first amplitude spectrum of each GC was calculated on the time interval [0, 60] billion years, the second - on [60, 120] billion years. Then the frequency drift parameters were calculated for each time series $x(t_n), y(t_n), z(t_n)$ using the formula (\ref{freq}) and the largest value of them was taken as the frequency drift parameter $\lg( \Delta f)$. In the case of the same fundamental frequencies $\Omega_i(t_1)=\Omega_i(t_2)$, we artificially assumed the frequency drift parameter to be equal to $-4$.

The results of calculating the frequency drift parameter and the nature of the GC motion - (R) or (C), which was determined from a joint analysis of the values of the frequency drift parameters and visual analysis of the amplitude spectra, are reflected in the seventh column of Table~1 (see also Fig.~10).
 It turned out that in our case, the division of orbits into regular and chaotic can be done using the threshold value of the parameter $\lg(\Delta f)=-2.26$. The smaller value corresponds to regular orbits, the larger value corresponds to chaotic ones, with the exception of two GCs: Terzan 3 and NGC 6316, for which $\lg(\Delta f)\approx -2$, but we classified them as GCs based on the results of a visual analysis of the amplitude spectra with regular orbits, although they show a weak degree of chaos.

Fig.~\ref{Lyap6} (top row) shows an example of spectra for regular (NGC 6266) and chaotic (NGC 6355) motion, where the amplitude spectrum of the first half of the time sequence is shown in red, and the second half in black. In the bottom row of Fig.~\ref{Lyap6} on the left panel there is a histogram of the distribution of the frequency drift parameter
$\lg(\Delta f)$, the right panel shows the diagram $"$frequency drift - HLE$"$, showing a good correlation between the values of the frequency drift parameter and HLE, the correlation coefficient is $K_c=$0.76 (see also Table~2). A graphical illustration of the method for all 45 ALs is presented in Fig.~10 in the lower horizontal row of panels. The correlation coefficients of classification results with other methods are given in Table~3. The smallest correlation ($K_c=$0.64) is observed with the results of calculating the HLE with renormalization, the largest (0.96) with the Poincar$\acute{e}$ section method.

 To further formalize the process of separating GCs with regular and chaotic dynamics, let us turn to the histogram of the distribution of the frequency drift parameter in Fig.~\ref{Lyap6}. In accordance with the histogram, we classify GCs with regular dynamics as those with a frequency drift parameter $\lg(\Delta f)< -3$, and GCs with average and weak chaotic dynamics if $-2.5 <\lg(\Delta f )< -1$, and to a GC with highly chaotic dynamics, if
$-0.5 <\lg(\Delta f)< 0$.

\subsection{Visual assessment of orbital regularity/chaoticity}
\begin{figure*}
{\begin{center}
               \includegraphics[width=0.17\textwidth,angle=-90]{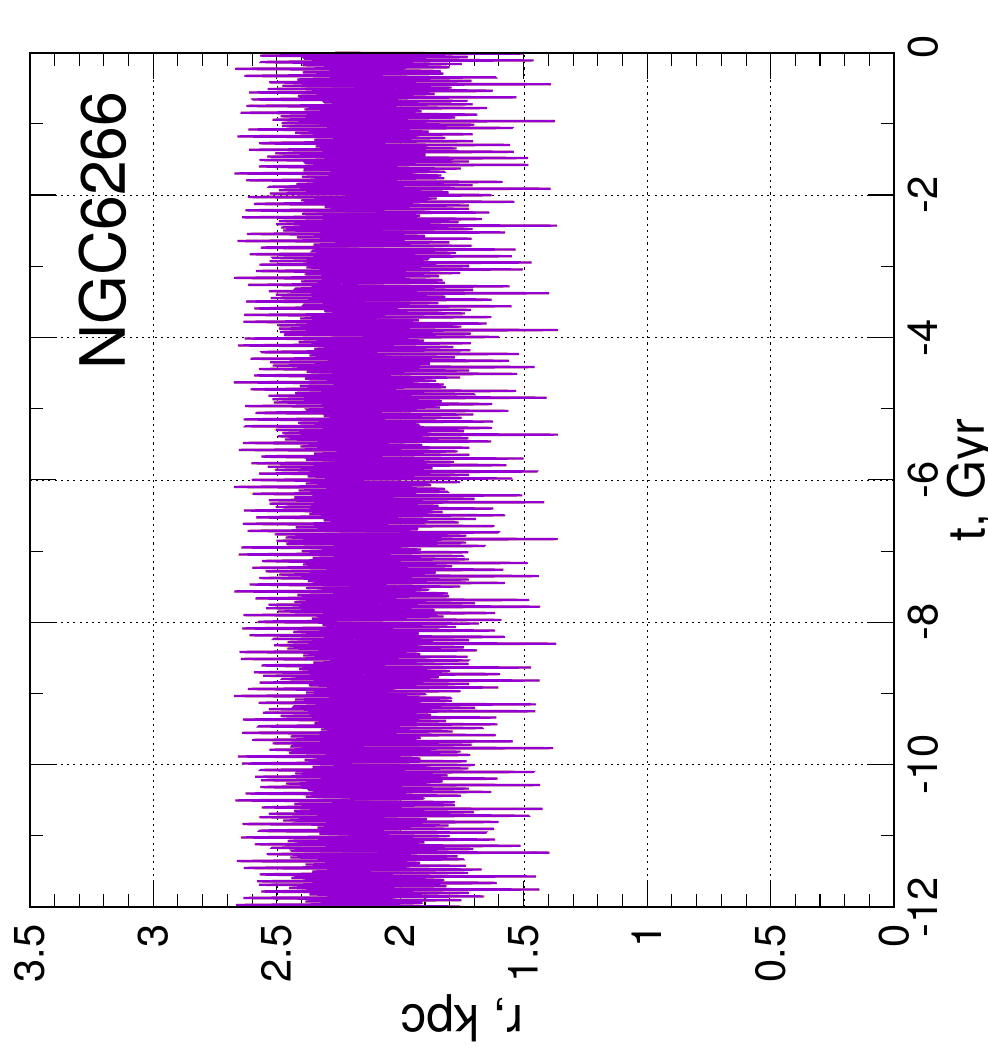}
               \includegraphics[width=0.17\textwidth,angle=-90]{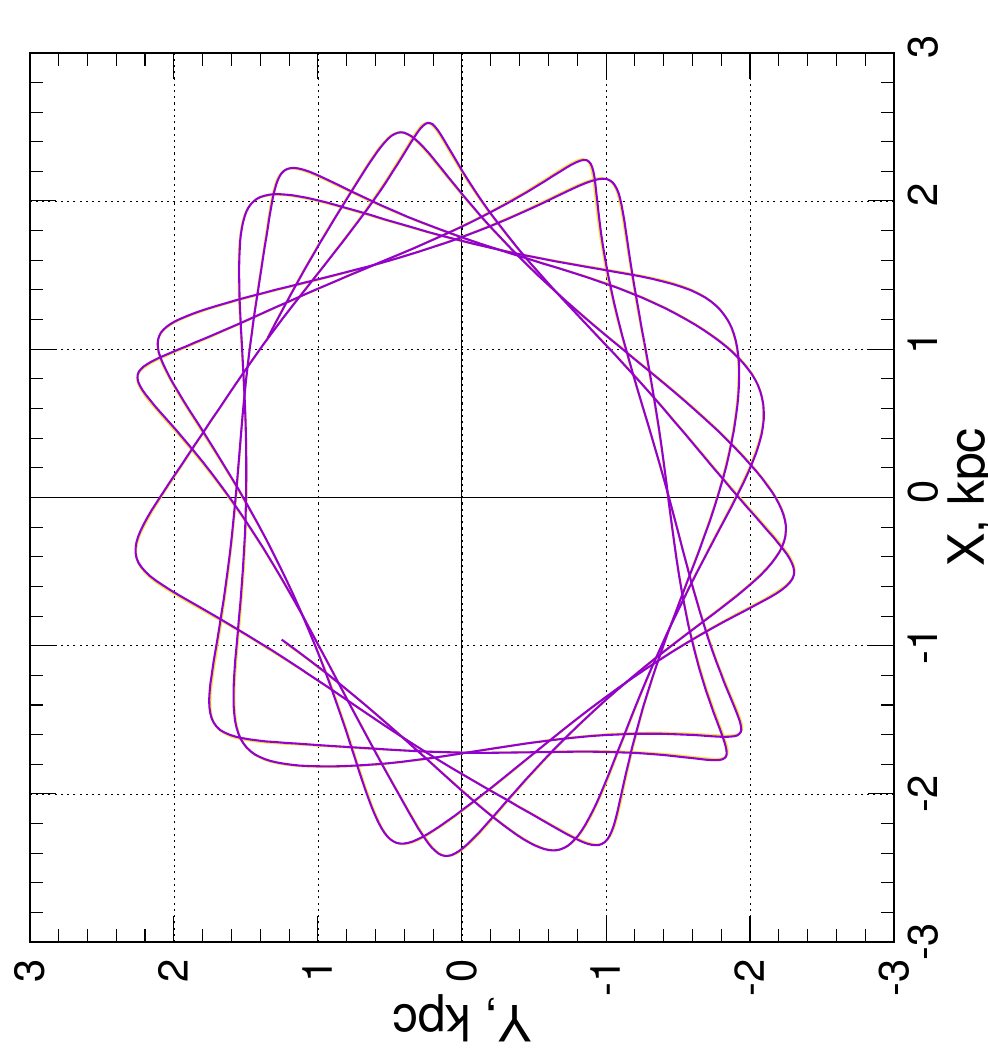}               \includegraphics[width=0.17\textwidth,angle=-90]{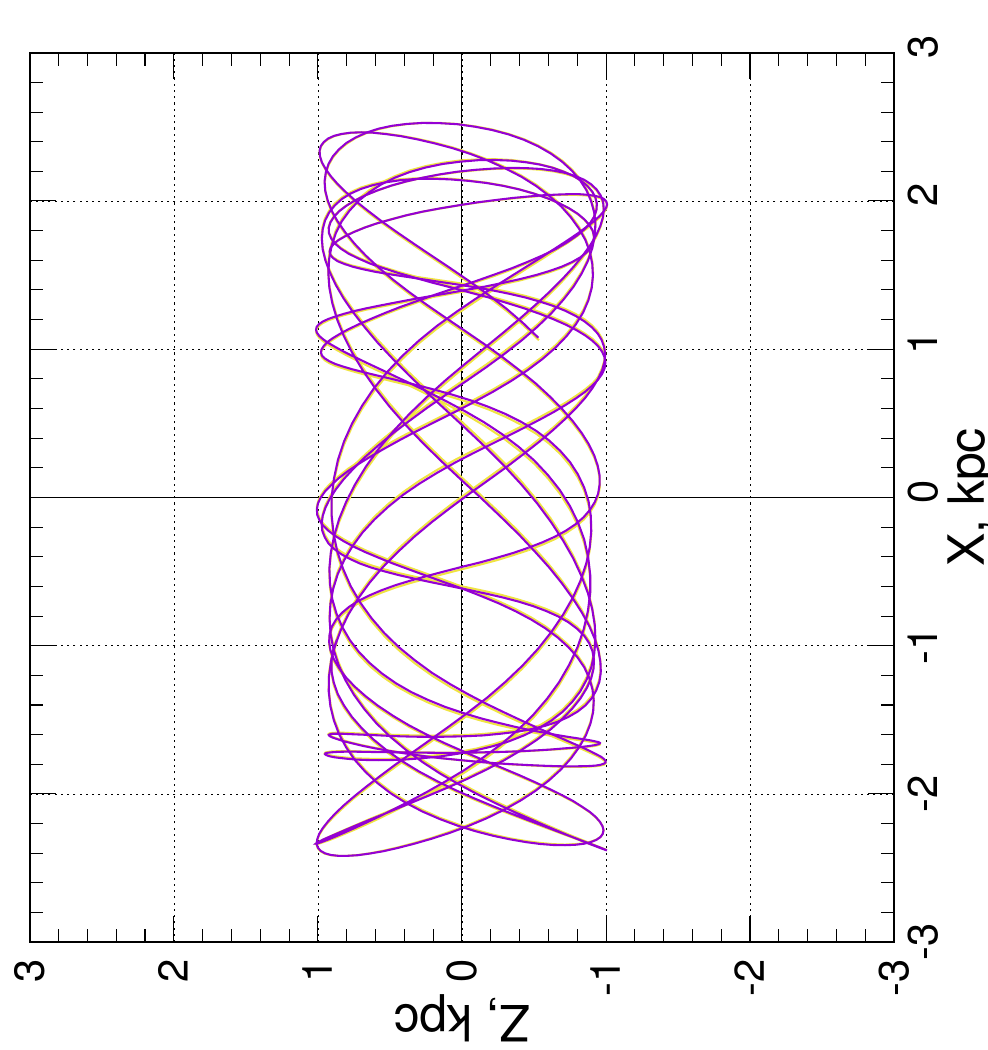}               \includegraphics[width=0.17\textwidth,angle=-90]{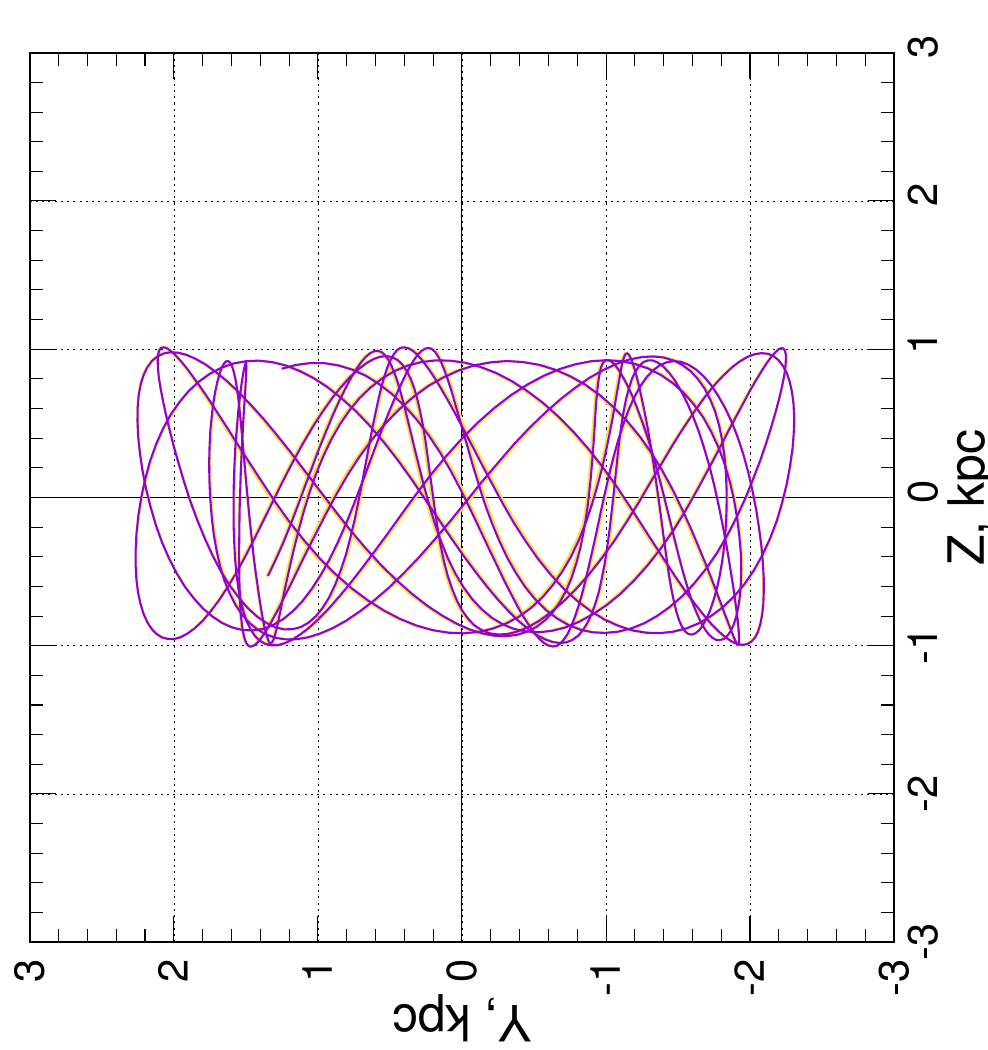} \

               \includegraphics[width=0.17\textwidth,angle=-90]{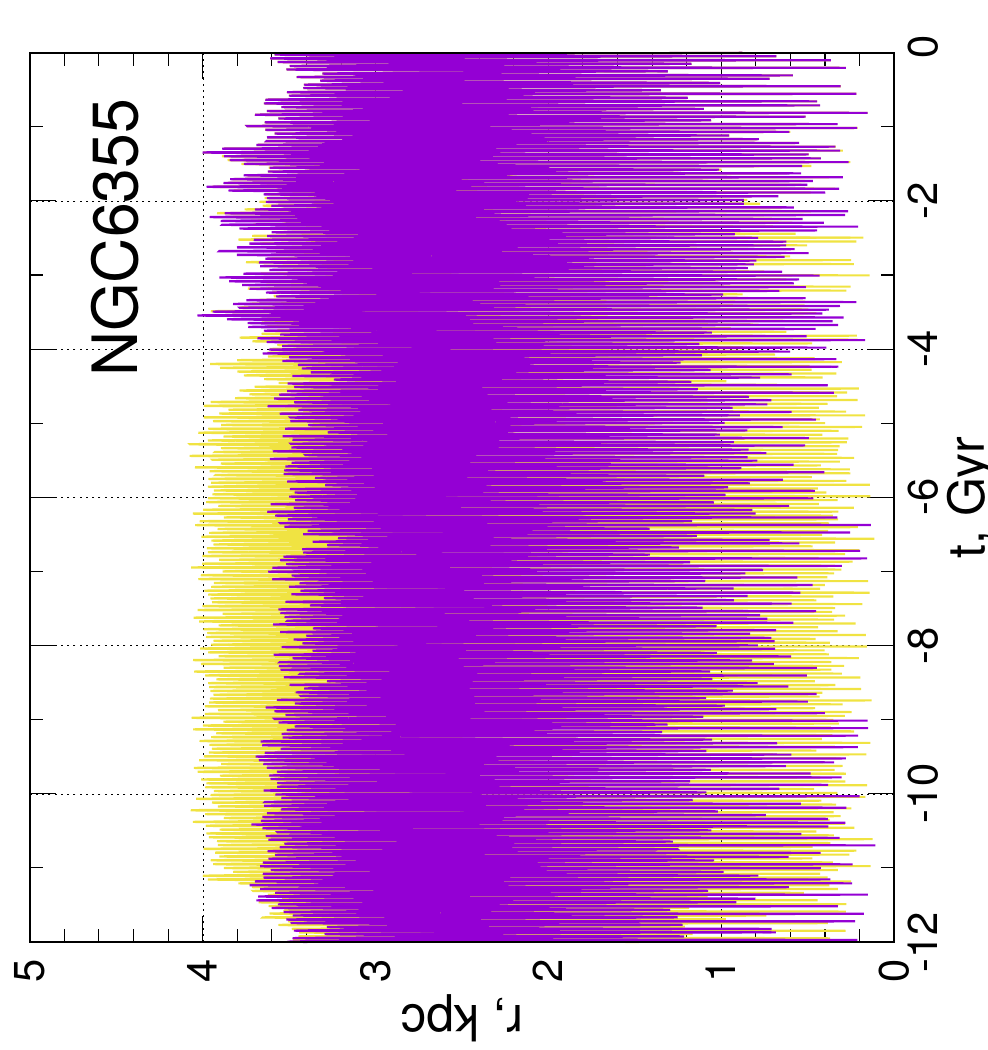}
               \includegraphics[width=0.17\textwidth,angle=-90]{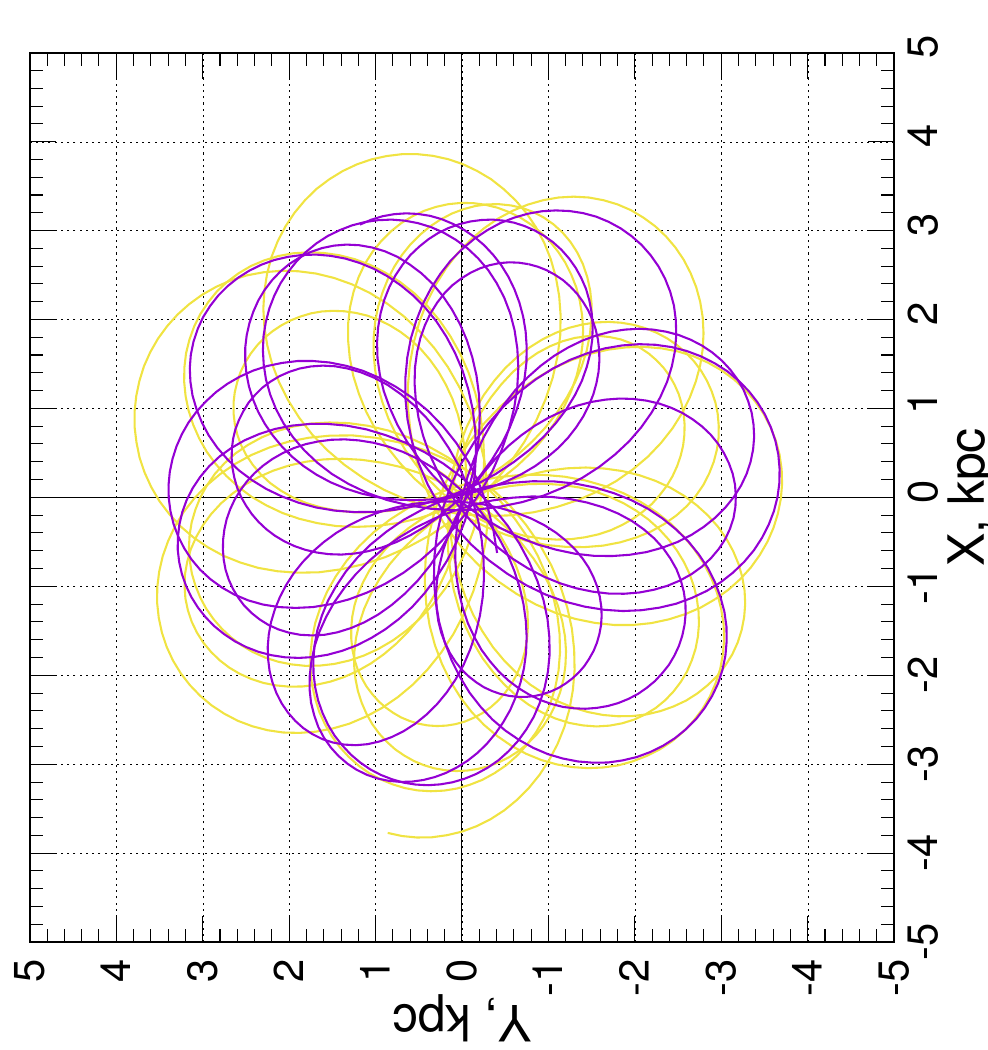}               \includegraphics[width=0.17\textwidth,angle=-90]{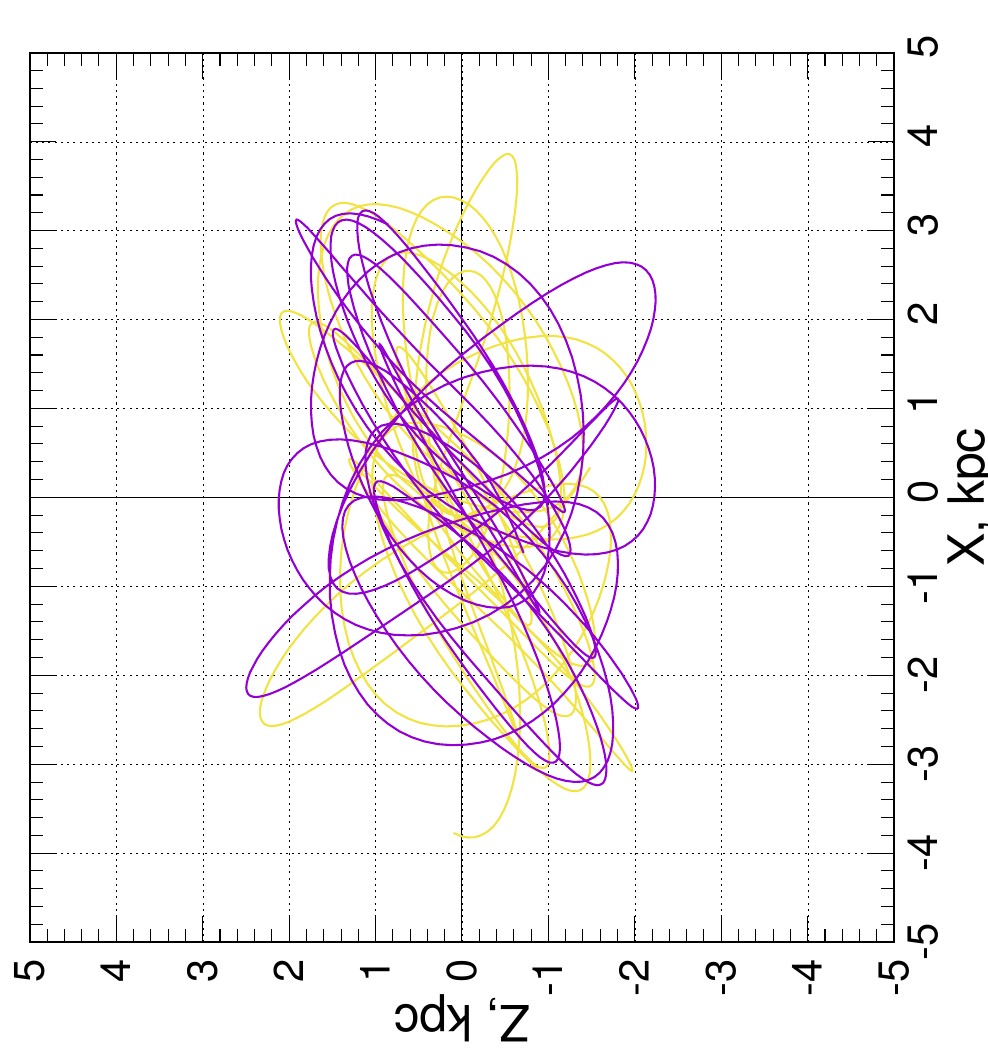}               \includegraphics[width=0.17\textwidth,angle=-90]{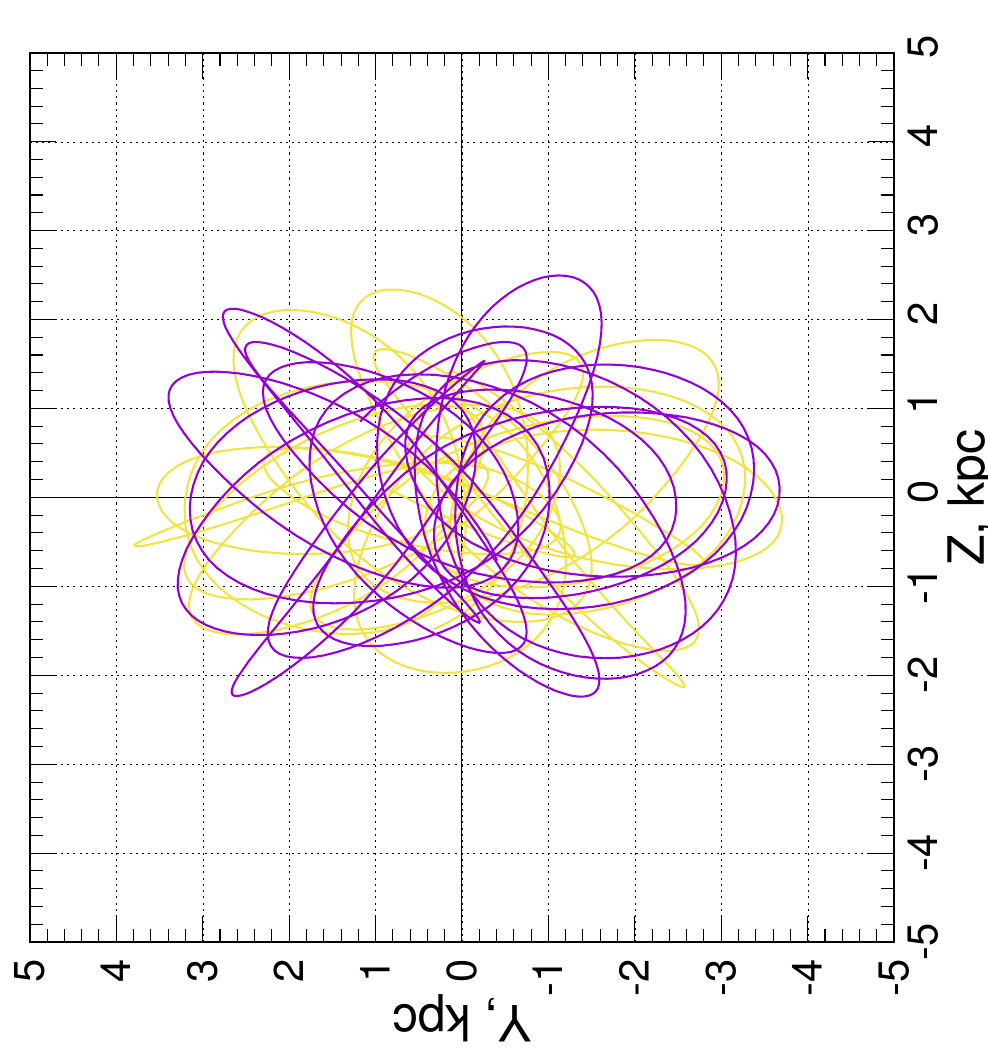} \

\bigskip

\caption{\small Orbits of globular clusters NGC 6266 (R) (top panels), NGC 6355 (C) (bottom panels). In the panels from left to right: radial orbital values as a function of time over the interval [0,-12] billion years; $X-Y$, $X-Z$, $Z-Y$ projections of orbits constructed in the rotating bar system over the time interval [-11,-12] billion years. The reference orbit is shown in yellow, the shadow orbit is shown in purple.}
\label{Lyap7}
\end{center}}
\end{figure*}

The most informative illustration of the discrepancy between the reference and shadow phase points is Fig.~10, which shows the reference and shadow orbits for each GC in the order (from left to right) as they are indicated in Table~1. The topmost panels show the radial values of the orbit as a function of time over the interval $[0,-12]$ billion years, comparable to both the age of the GC and the Universe; below vertically (from the second to the fourth horizontal row) are $X-Y$, $X-Z$, $Z-Y$ projections of orbits, respectively, constructed in the rotating bar system on the time interval $[-11,-12]$ billion years. In all of these graphs, the reference orbits are shown in yellow, and the shadow orbits in purple. You can see that many objects on the graphs only have purple color. This means that the shadow orbit practically coincides with the reference one (yellow lines are covered with purple ones). Such objects include GCs with regular orbits. On the graphs of GCs with chaotic orbits, both purple and yellow lines are visible, which allows us to qualitatively judge the degree of chaotic orbits. The results of the visual assessment - (R) or (C) - are given in the eighth column of Table~1. As an example, Fig.~\ref{Lyap7} shows an illustration for the GC NGC 6266 with a regular orbit and NGC 6355 with a chaotic orbit.

\subsection{Estimating the probability of orbital regularity using the $"$voting$"$ method}
Thus, above we analyzed the regularity/chaosity of the orbits of 45 GCs in the central region of the Milky Way, where the influence of the bar is most strongly felt, using six methods. The classification results for GCs with regular (R) and chaotic (C) orbits obtained by each method are summarized in Table~1. As can be seen from the table, classification results obtained by different methods may differ. For a final assessment of the degree of regularity/chaosity of the orbit, we introduced a new characteristic - the probability of orbit regularity $P(R)$, which is calculated as the ratio of the number of methods that characterize the orbit as regular to the total number of methods used, in this case equal to six. The probability of chaos is calculated as $P(C)=1-P(R)$. We call this method of calculating probability the $"$voting$"$ method. The obtained values of $P(R)$ along with the final solution: (R) or (C) are given in the ninth column of Table~1. Decision (R) was made if $P(R)>0.5$, decision (C) if $P(R)\leq 0.5$).

\begin{figure*}
{\begin{center}
               \includegraphics[width=0.3\textwidth,angle=-90]{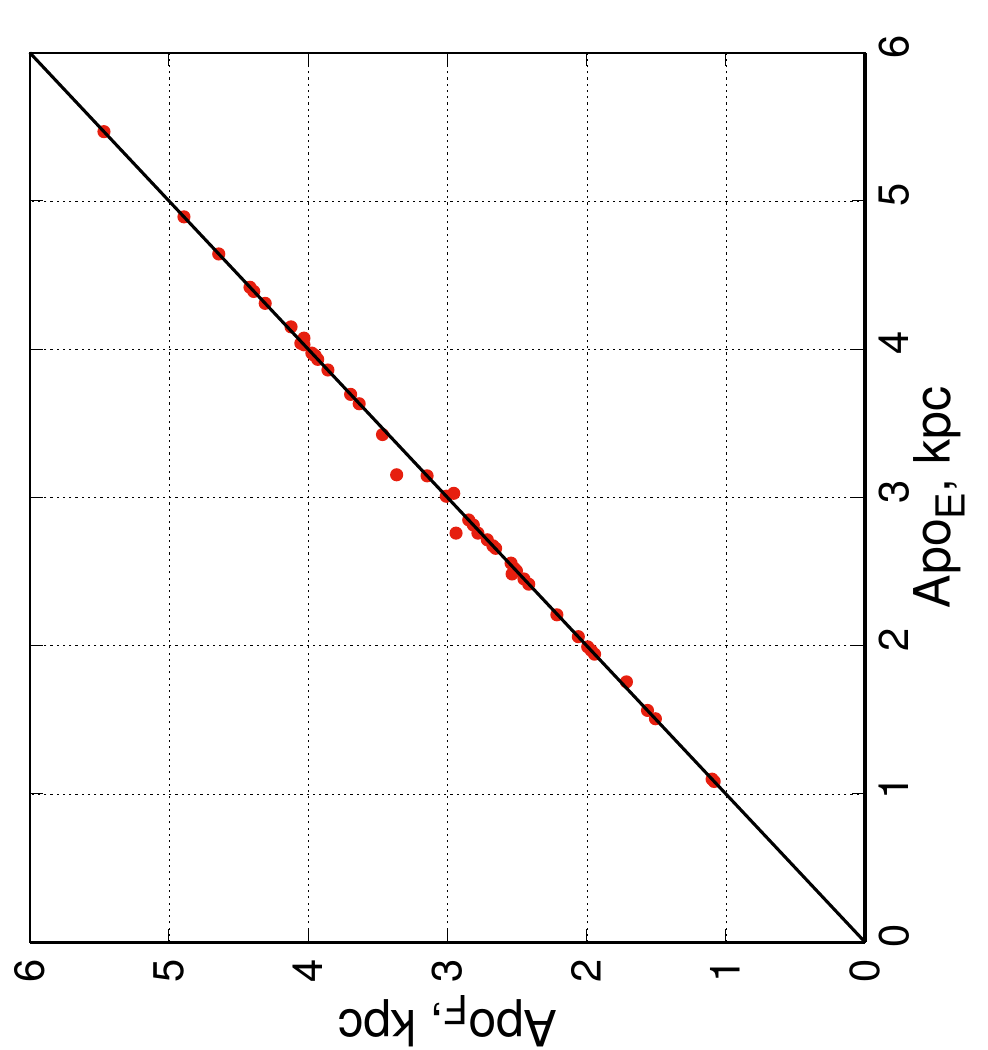}
               \includegraphics[width=0.3\textwidth,angle=-90]{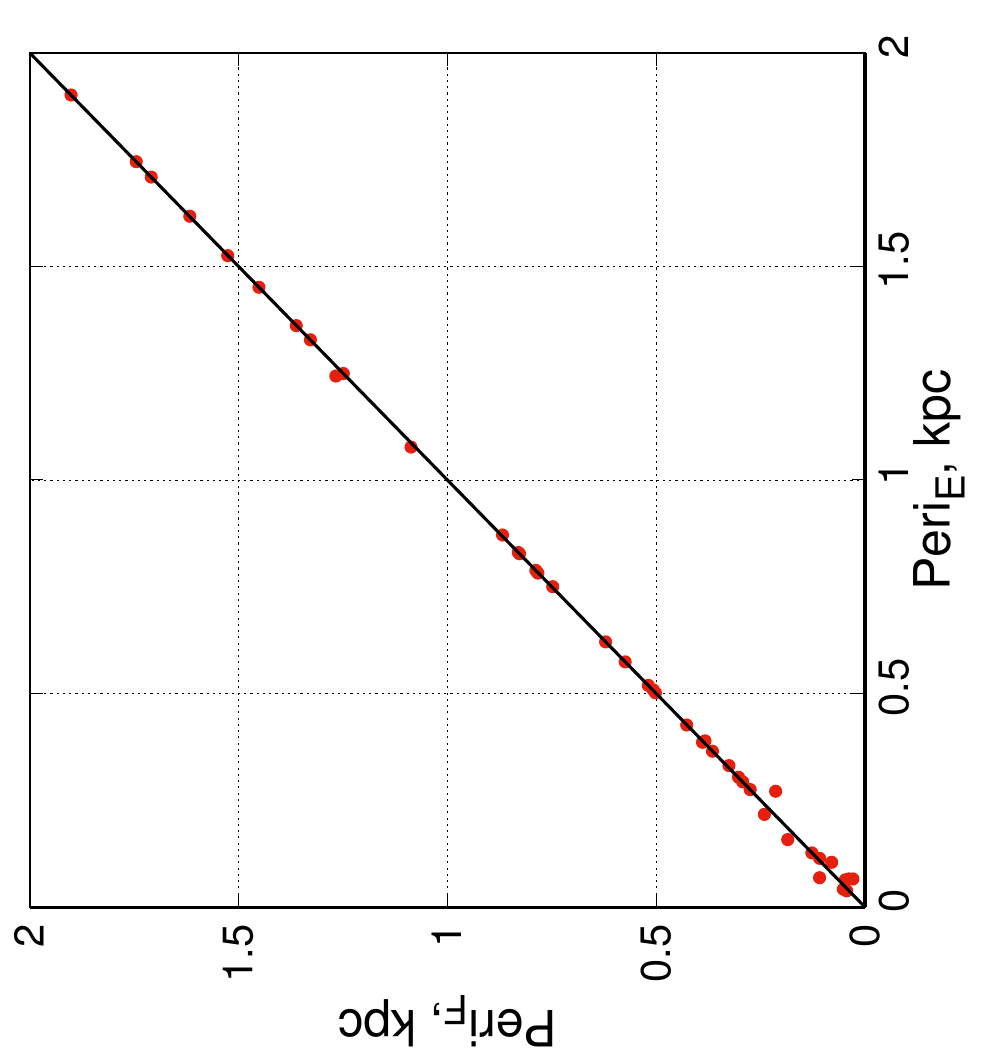}
               \includegraphics[width=0.3\textwidth,angle=-90]{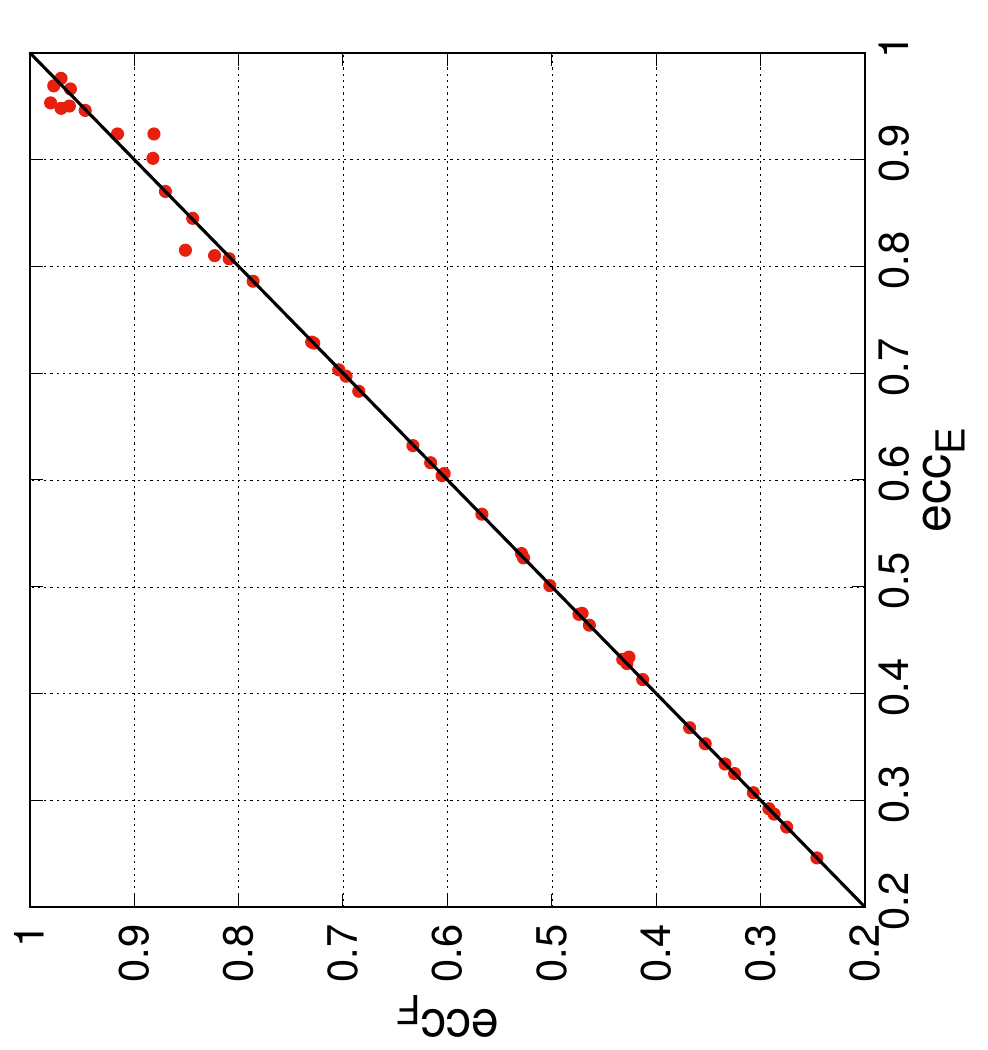}\

\bigskip

\caption{\small Comparison of the parameters of the reference (E) and shadow (F) orbits for 45 GCs (left panel - apocentric distance, middle panel - pericentric distance, right panel - orbital eccentricity).}
\label{Lyap71}
\end{center}}
\end{figure*}

\begin{figure*}
{\begin{center}
               \includegraphics[width=0.3\textwidth,angle=-90]{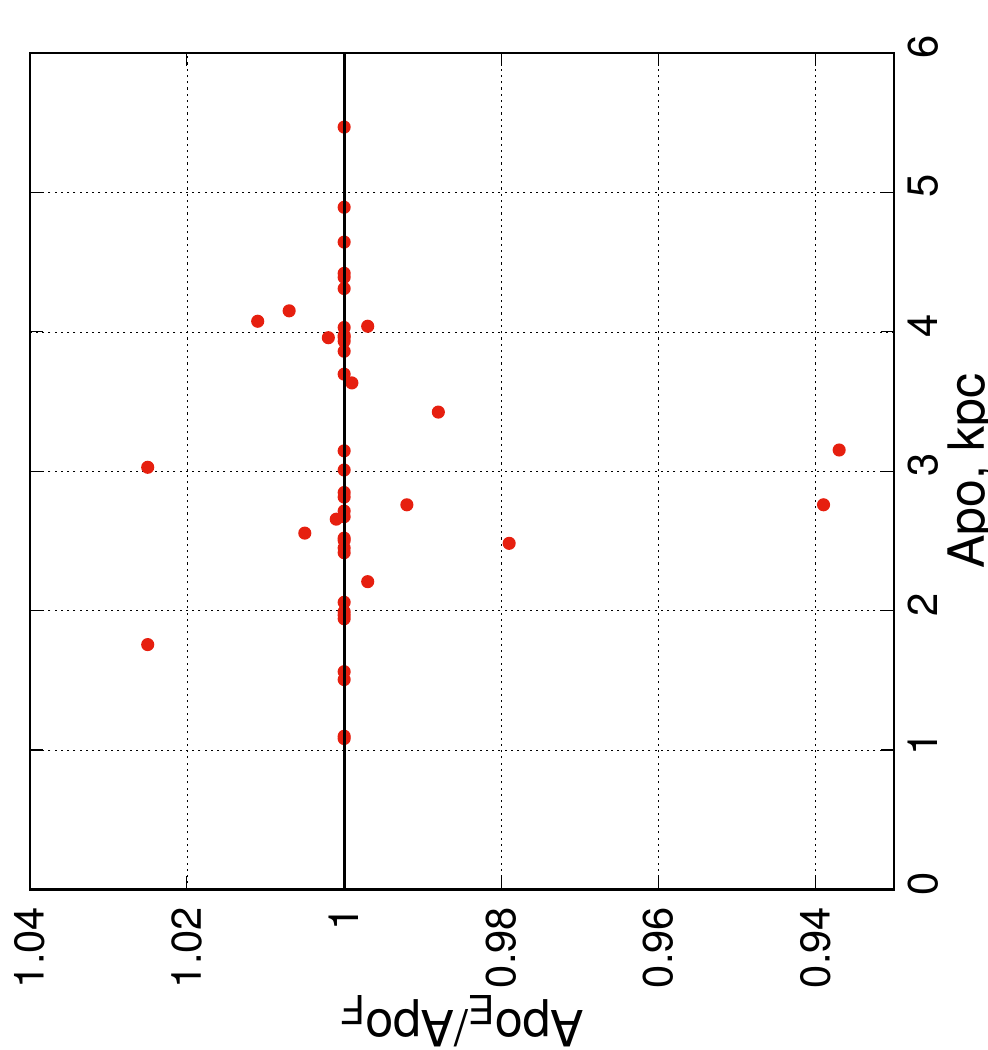}
               \includegraphics[width=0.3\textwidth,angle=-90]{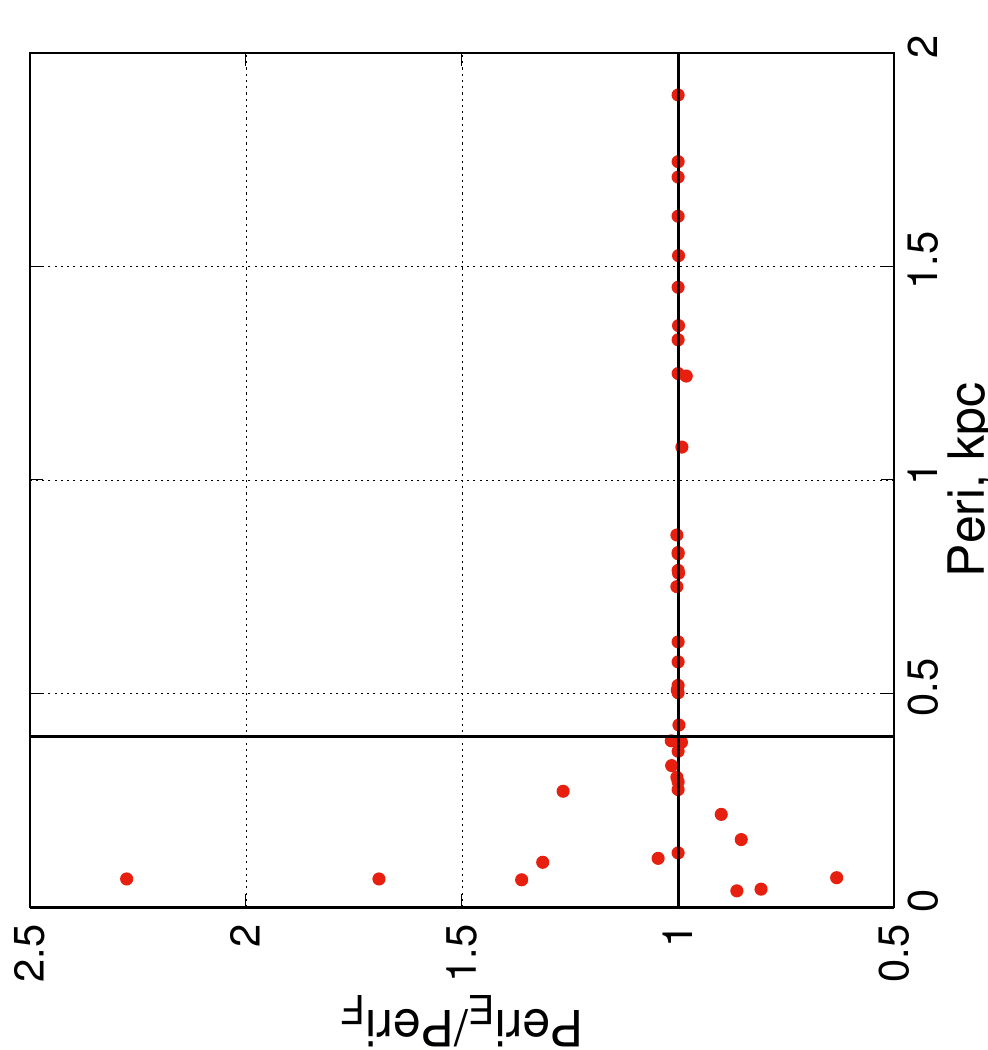}
               \includegraphics[width=0.3\textwidth,angle=-90]{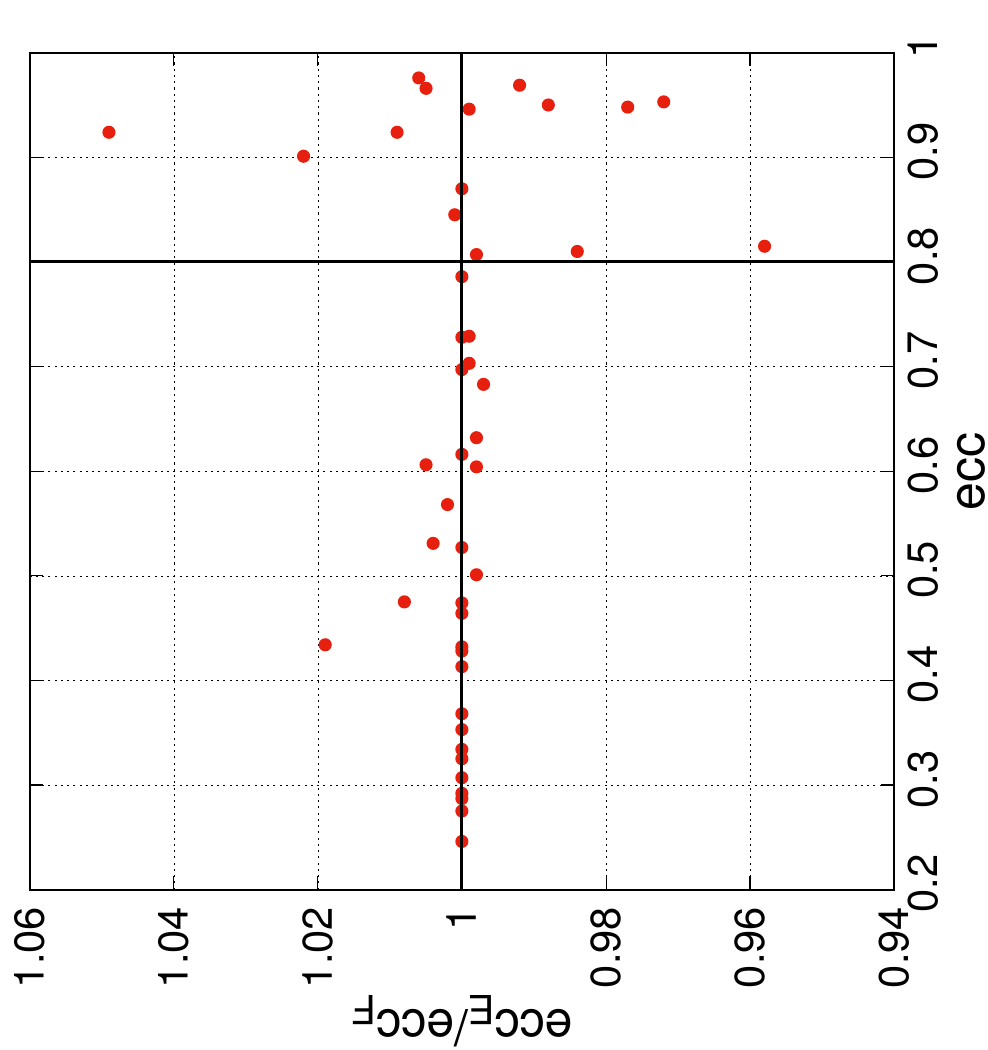}\

\bigskip

\caption{\small Relative deviations of the parameters of the shadow (F) from the reference (E) orbits of 45 GC (left panel - apocentric distance, middle panel - pericentric distance, right panel - orbital eccentricity).}
\label{Lyap72}
\end{center}}
\end{figure*}

\section{Comparison of classification results for GCs with regular and chaotic dynamics}
The results of the classification of 45 GCs on the basis of regularity/chaoticity of orbits, obtained in the previous section by seven different methods, are summarized in Table~1. The seventh method is to determine the probability of the final decision $P(R)$ using the $"$voting$"$ method. A graphic illustration of the methods for analyzing the regularity of orbits in relation to each GC is given in Fig.~10.

In order to compare the results of GC classification based on the data in Table~1, two correlation matrices were calculated. The first one is based on the numerical results of three methods: MEGNO, HLE and frequency method. The second one is based on the classification results obtained by six methods, when the solution (R) is assigned the number $"1"$, and the solution (C) is assigned the number $"0"$, as well as the probability values $P(R)$ (seventh method). The resulting correlation matrices 1 and 2 are presented in the form of tables 2 and 3, respectively.

Let's consider correlation matrix 1. The smallest correlation is observed between MEGNO and the frequency method (correlation coefficient $K_c=0.65$), the largest - between HLE and the frequency method ($K_c=0.76$). Let's move on to correlation matrix 2, compiled according to the classification results: (R) or (C). We see that the probabilistic method gives the lowest correlation with HLE ($K_c=0.60$) and the highest with MEGNO ($K_c=0.95$) and the visual method ($K_c=0.96$). MEGNO gives the lowest correlation with HLE ($K_c=0.65$) and the highest with the probabilistic method ($K_c=0.95$). HLE gives the lowest correlation with the visual method ($K_c=0.55$) and the highest with MEGNO ($K_c=0.65$). The Poincar$\acute{e}$ section method gives the lowest correlation with the HLE ($K_c=0.60$) and the highest with the frequency and visual methods ($K_c=0.96$). The frequency method gives the lowest correlation with the HLE ($K_c=0.64$) and the highest with the Poincar$\acute{e}$ section method ($K_c=0.96$). The visual method gives the lowest correlation with the HLE ($K_c=0.55$) and the highest with the probabilistic method and the Poincar$\acute{e}$ section method ($K_c=0.96$). The probabilistic decision-making method based on the principle of $"$voting$"$~ correlates with HLE with $K_c=0.74$, and with all other methods with $K_c=0.94\div 0.96$. As a result, we can conclude that the lowest correlation with other methods is observed in the HLE, and the highest in the visual method. The Poincar$\acute{e}$ section method for decision making is close to the frequency method, both of these methods correlate well with each other and other methods. However, it is necessary to note the difference, although not very large, between the correlation coefficients regarding MEGNO, HLE and the frequency method obtained from numerical data (Table~2) and from decision-making results (Table~3).

In general, we can conclude that all methods correlate well with each other. As the final decision on the classification of GCs with regular (R) and chaotic (C) dynamics, we will accept the classification results based on calculating the probabilities $P(R)$, since it showed the greatest correlation with all other methods. As a result, the list of GCs with regular dynamics included 24 objects (NGC 6266, Terzan 4, Liller 1, NGC 6380, Terzan 1, Terzan 5, Terzan 6, Terzan 9, NGC 6522, NGC 6528, NGC 6624, NGC 6637, NGC 6717, NGC 6723, Terzan 3, Pismis 26, NGC 6569, E456-78, NGC 6540, Djorg 2, NGC 6171, NGC 6316, NGC 6539, NGC 6553), and the list of GCs with chaotic dynamics includes 21 (NGC 6144 , E452-11, NGC 6273, NGC 6293, NGC 6342, NGC 6355, Terzan 2, BH 229, NGC 6401, PAL 6, NGC 6440, NGC 6453, NGC 6558, NGC 6626, NGC 6642, NGC 6256, NGC 6256, NGC 6256, NGC 6256, NGC 6304, NGC 6325, NGC 6388, NGC 6652). Note that the final lists differ from the initial ones obtained by the probabilistic method (see Section 1.1), and differ only in two objects: NGC 6304 and NGC 6388, which migrated from the first list of objects with regular motion to the list of objects with chaotic motion.

To once again verify the reason for the chaotic movement of the GCs, we also carried out
comparison of such parameters as apocentric distance, pericentric distance and eccentricity of the reference and shadow orbits for all 45 GCs 12 billion years ago (Fig.~\ref{Lyap71}). It is obvious that GCs with regular orbits should have deviations from the diagonal line of comparison close to zero (which also follows from a visual analysis of the orbits). As can be seen from Fig.~\ref{Lyap71}, GCs with chaotic orbits already give noticeable deviations from the comparison line, especially in the region of small values of the pericentric distance ($<0.4$ kpc) and large values of eccentricity ($>0.8$) due to the most strong influence of the central bar on the orbital motion of the GC precisely in the region closest to the center of the Galaxy. Apocentric distances deviate maximum from the comparison line in the region of 2.5-3.5 kpc. The degree of deviations of orbital parameters is especially clearly visible in Fig.~\ref{Lyap72}, which shows the relative deviations of the parameters of the shadow orbit from the reference orbit. It should be noted that the relative deviations of the apocentric distances are small. The greatest relative deviations are experienced by pericentric distances.

Note that the correlation between the deviations of pericentric distances and eccentricities from the corresponding comparison lines is quite large and amounts to 0.85. These two features can serve as unique indicators of the chaotic nature of the GC orbits.
Thus, globular clusters with recentric distances $<0.4$ kpc and relative deviations of recentric distances $|Peri_E/Peri_F-1|>0.01$, where the index $E$ refers to the reference orbit, and the index $F$ to the shadow orbit, are E452-11, NGC 6273, NGC 6355, Terzan 2, BH 229, NGC 6401, Pal 6, NGC 6453, NGC 6558, NGC 6626, NGC 6638, NGC 6642, NGC 6256, NGC 6652, and GCs with large eccentricities ($> 0.8$) -- E452-11, NGC 6273, NGC 6293, NGC 6355, Terzan 2, BH 229, NGC 6401, Pal 6, NGC 6440, NGC 6453, NGC 6558, NGC 6626, NGC 6638, NGC 6642, NGC 6652. The intersection of these two sets is 13 objects. All of the listed GCs are included in our final list of GCs with chaotic dynamics.

Thus, we can conclude that the list of GCs with chaotic dynamics includes mainly globular clusters with elongated radial orbits with small pericentric distances ($<0.4$ kpc) and large eccentricities ($>0.8$). In general, our result is consistent with the conclusions of [11] that chaotic orbits occupy the bar region.

\section*{CONCLUSIONS}

1. For the first time, the regularity/chaoticity of the orbital dynamics of a sample of 45 globular clusters in the central region of the Milky Way with a radius of 3.5 kpc was studied in a non-axisymmetric gravitational potential [2,3] with an elongated rotating bar. We used a bar model in the form of a triaxial ellipsoid with a mass of $10^{10} M_\odot$, a semi-major axis length of 5 kpc, an inclination angle to the galactic axis $X$ of 25$^o$, and a rotation velocity of 40 km s$^{-1}$ kpc$^{-1}$. To construct the orbits, we used the most accurate astrometric data to date from the Gaia satellite (EDR3) [7], as well as new refined average distances to globular clusters [8].

2. As a result of the initial calculation of the Highest Lyapunov Exponents in a direct way (i.e., without renormalizing the shadow trajectory of the GC), bimodality was discovered in the histogram of the distribution of approximations of Lyapunov exponents. Using a method based on approximating a histogram with two Gaussian probability distributions, the GC sample was divided into objects with regular and chaotic dynamics. An explanation for the detected bimodality is given. The proposed method is called as a $"$probabilistic$"$ one.

3. In addition to the probabilistic method, several other well-known methods were used to analyze GCs for regularity/chaoticity, namely, the calculation of the HLE with renormalization of the shadow orbit, MEGNO, the Poincar$\acute{e}$ section method, the frequency method based on estimating the fundamental frequency drift parameter, and also a method based on visual assessment of orbits over long time intervals comparable to the age of the Universe. By calculating correlation matrices, it is shown that all methods used correlate well with each other. The final decision on classifying GCs as objects with regular or chaotic dynamics was made according to the $"$voting$"$ principle, taking into account the results of the analysis of all methods used. As a result, two lists of globular clusters were formed.

4. The list of GCs with regular dynamics includes 24 objects: NGC 6266, Terzan 4, Liller 1, NGC 6380, Terzan 1, Terzan 5, Terzan 6, Terzan 9, NGC 6522, NGC 6528, NGC 6624, NGC 6637, NGC 6717 , NGC 6723, Terzan 3, Pismis 26, NGC 6569, E456-78, NGC 6540, Djorg 2, NGC 6171, NGC 6316, NGC 6539, NGC 6553.

5. The list of GCs with chaotic dynamics includes 21 objects: NGC 6144, E452-11, NGC 6273, NGC 6293, NGC 6342, NGC 6355, Terzan 2, BH 229, NGC 6401, Pal 6, NGC 6440, NGC 6453, NGC 6558, NGC 6626, NGC 6638, NGC 6642, NGC 6256, NGC 6304, NGC 6325, NGC 6388, NGC 6652.

6. As shown by the analysis of the parameters of the reference and shadow orbits at time intervals comparable to the age of the Universe, the list of GCs with chaotic dynamics includes predominantly globular clusters with elongated radial orbits with apocentric distances in the region of $2.5\div 3.5$ kpc, with small pericentric distances ( $<0.4$ kpc) and large eccentricities ($>0.8$) (the correlation between the last orbital parameters was 0.85), which is explained by the influence of the rotating bar on the dynamics of such GCs to the greatest extent, ultimately leading to chaotic motion.

\section*{ACKNOWLEDGMENTS}
The authors are grateful to the reviewer for a number of useful comments that helped improve the article. Special thanks to A.V. Melnikov for discussing issues related to the correct calculation of the HLE.

\newpage

\begin{figure*}
{\begin{center}

\includegraphics[width=1.0\textwidth,angle=0]{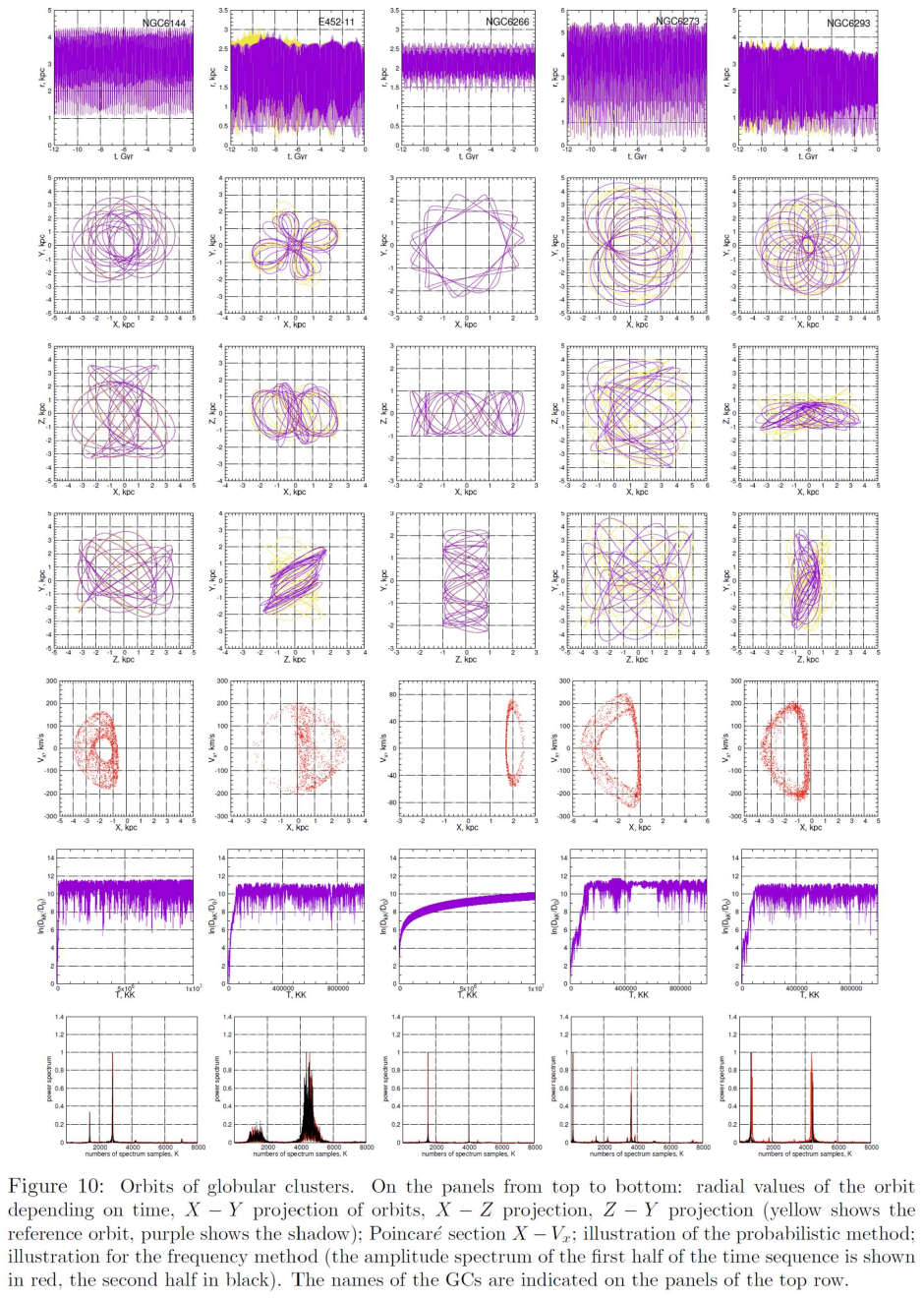}

\label{Lyap91}
\end{center}}
\end{figure*}

\newpage
\begin{figure*}
{\begin{center}

\includegraphics[width=1.0\textwidth,angle=0]{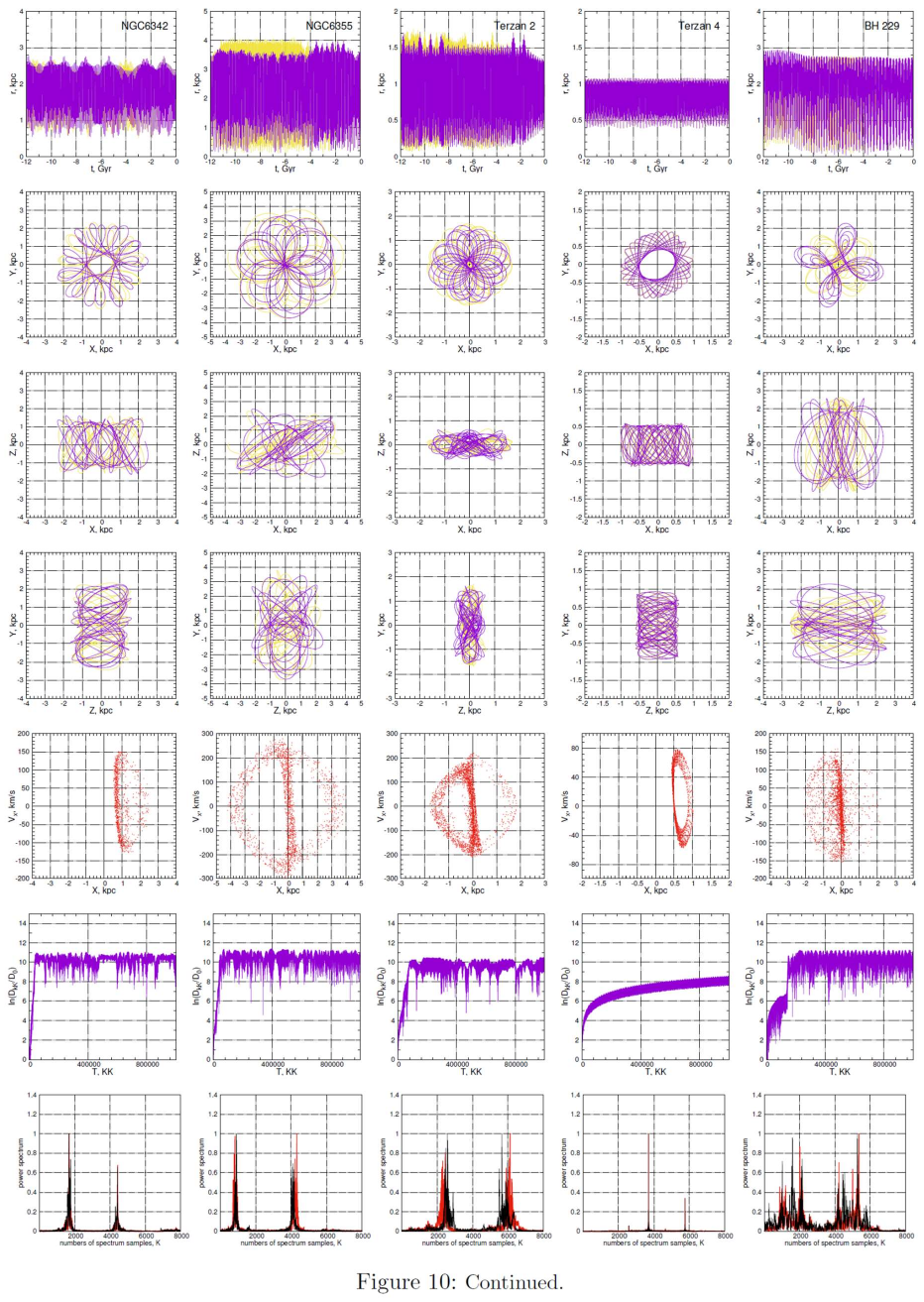}

\label{Lyap92}
\end{center}}
\end{figure*}

\newpage
\begin{figure*}
{\begin{center}

\includegraphics[width=1.0\textwidth,angle=0]{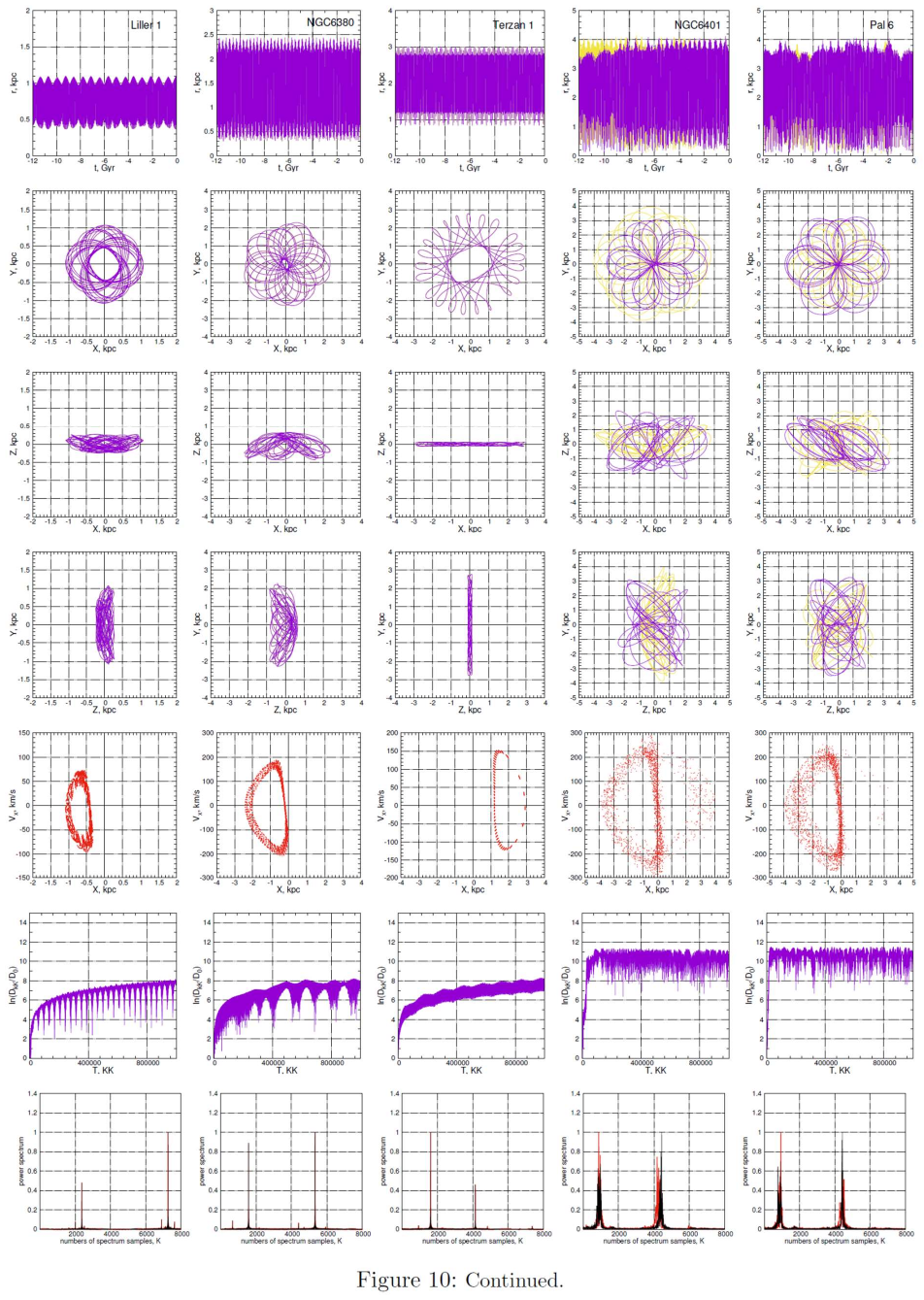}

\label{Lyap93}
\end{center}}
\end{figure*}

\newpage
\begin{figure*}
{\begin{center}

\includegraphics[width=1.0\textwidth,angle=0]{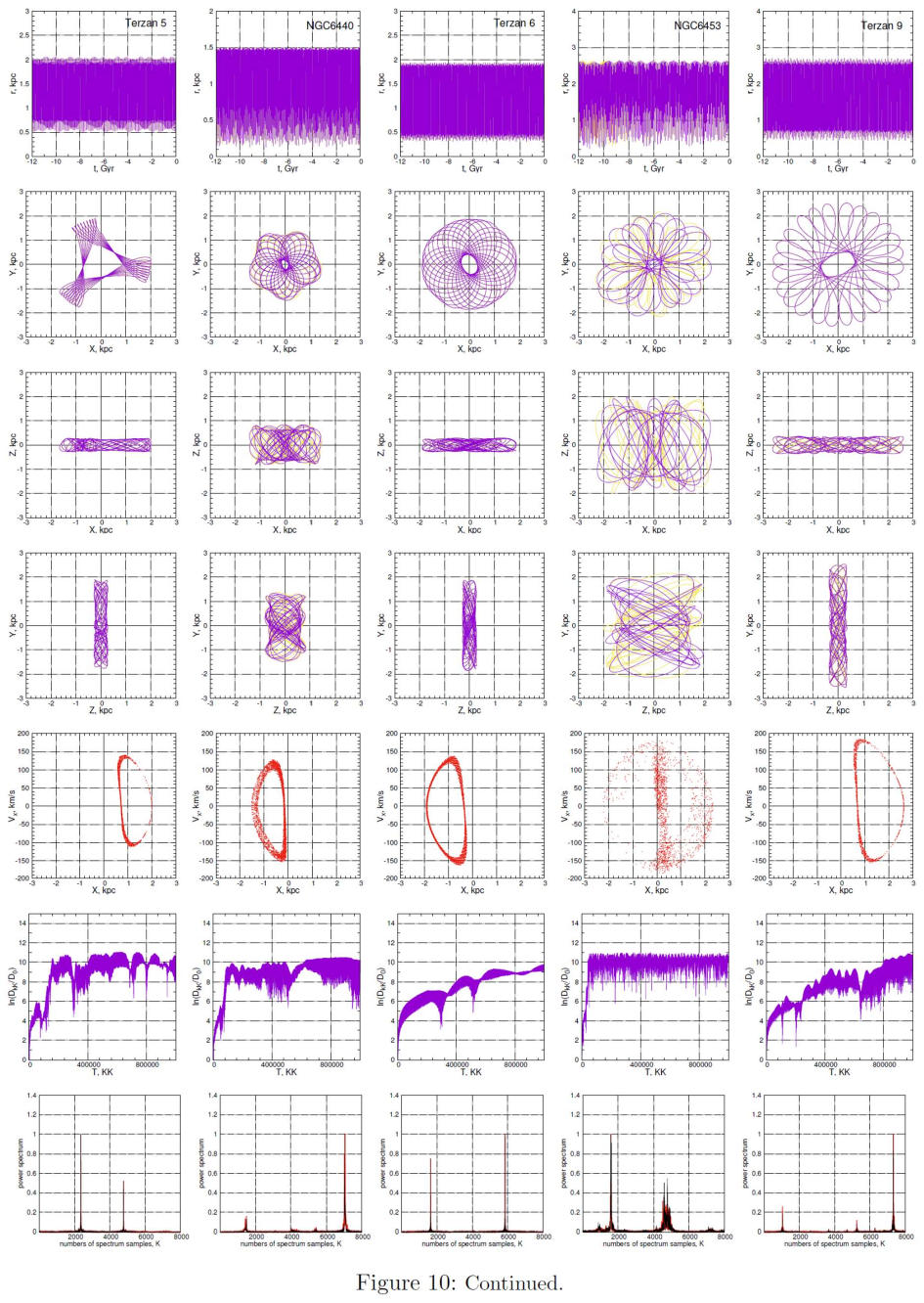}

\label{Lyap94}
\end{center}}
\end{figure*}

\newpage
\begin{figure*}
{\begin{center}

\includegraphics[width=1.0\textwidth,angle=0]{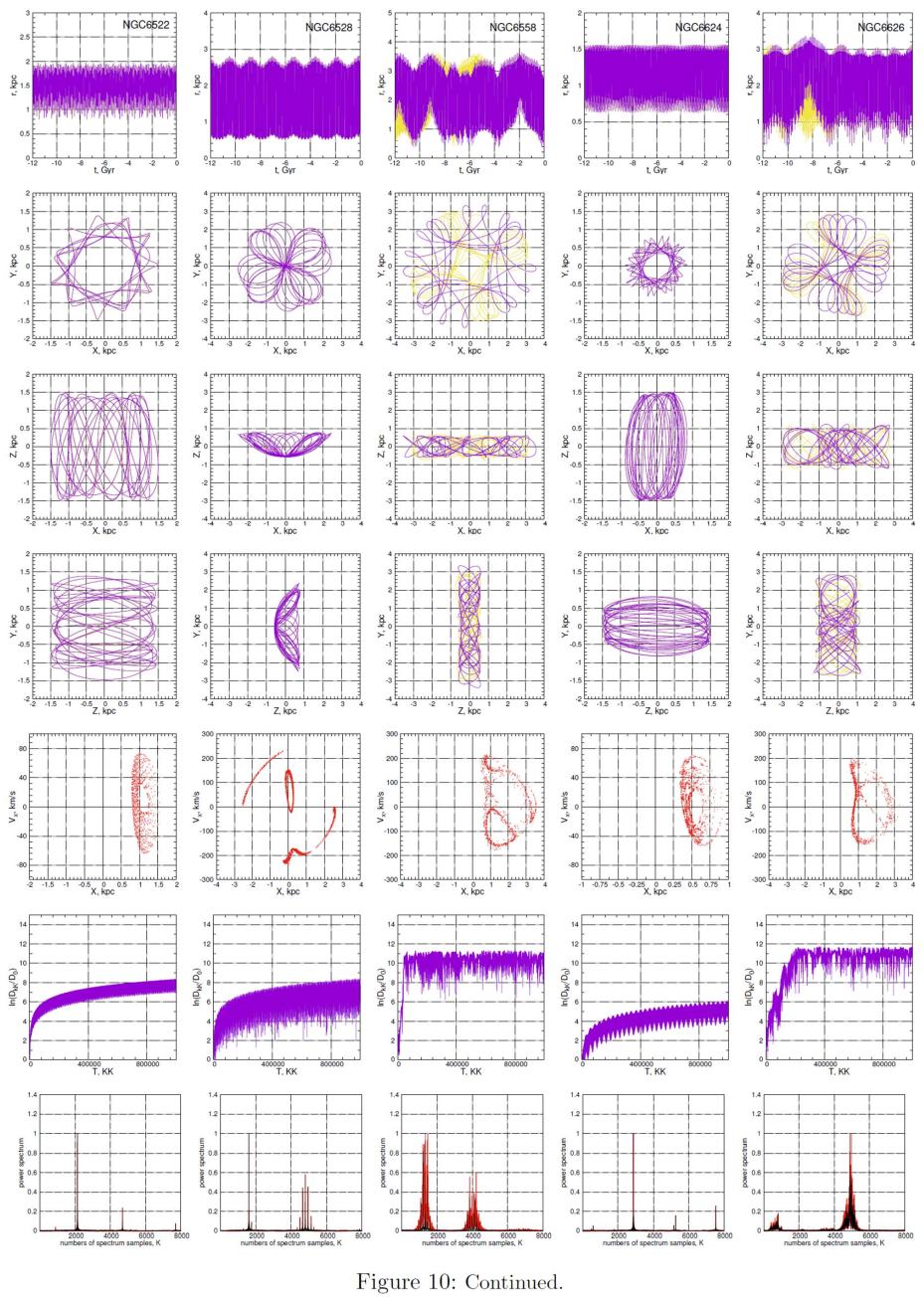}

\label{Lyap95}
\end{center}}
\end{figure*}

\newpage
\begin{figure*}
{\begin{center}

\includegraphics[width=1.0\textwidth,angle=0]{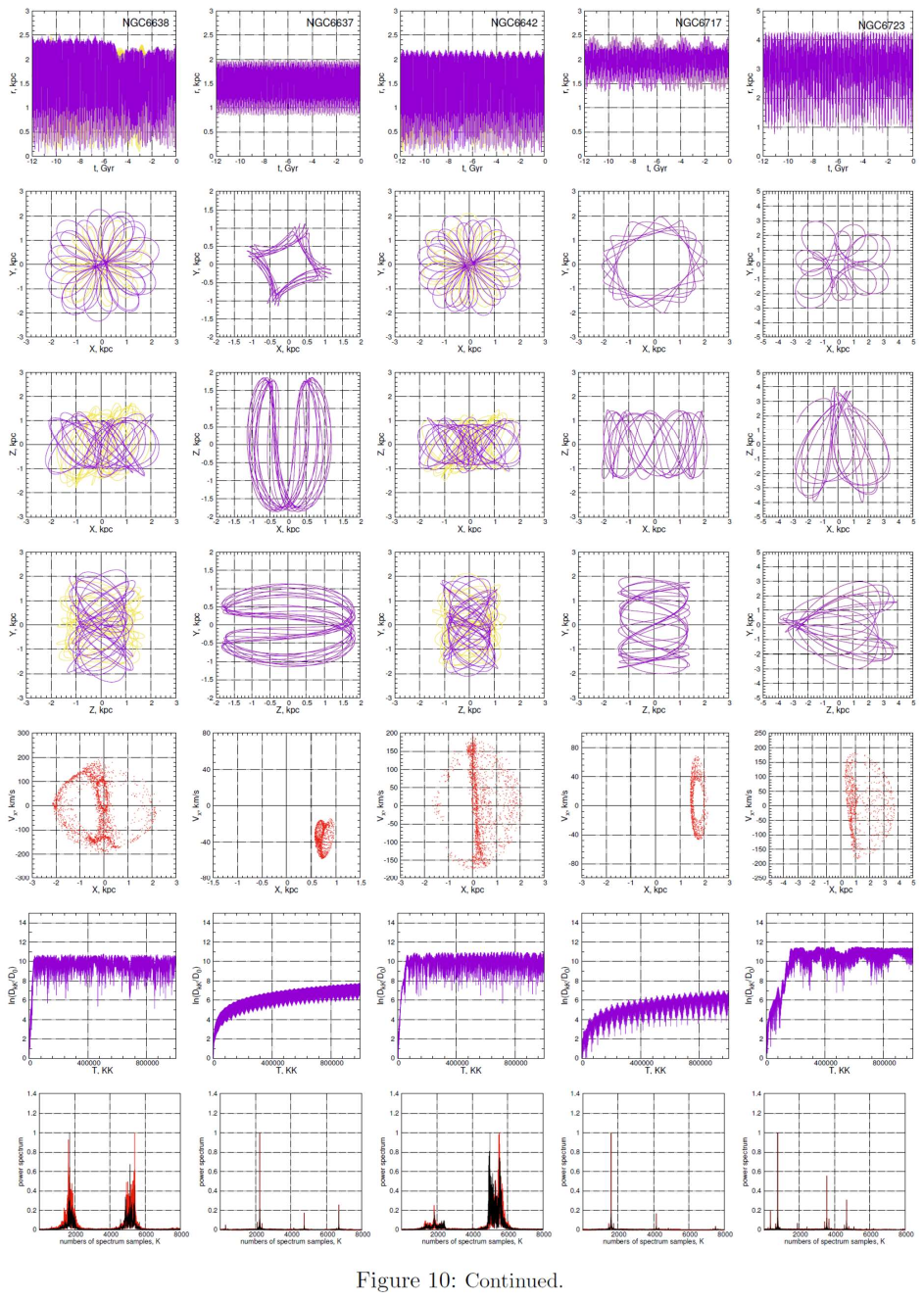}

\label{Lyap96}
\end{center}}
\end{figure*}

\newpage
\begin{figure*}
{\begin{center}

\includegraphics[width=1.0\textwidth,angle=0]{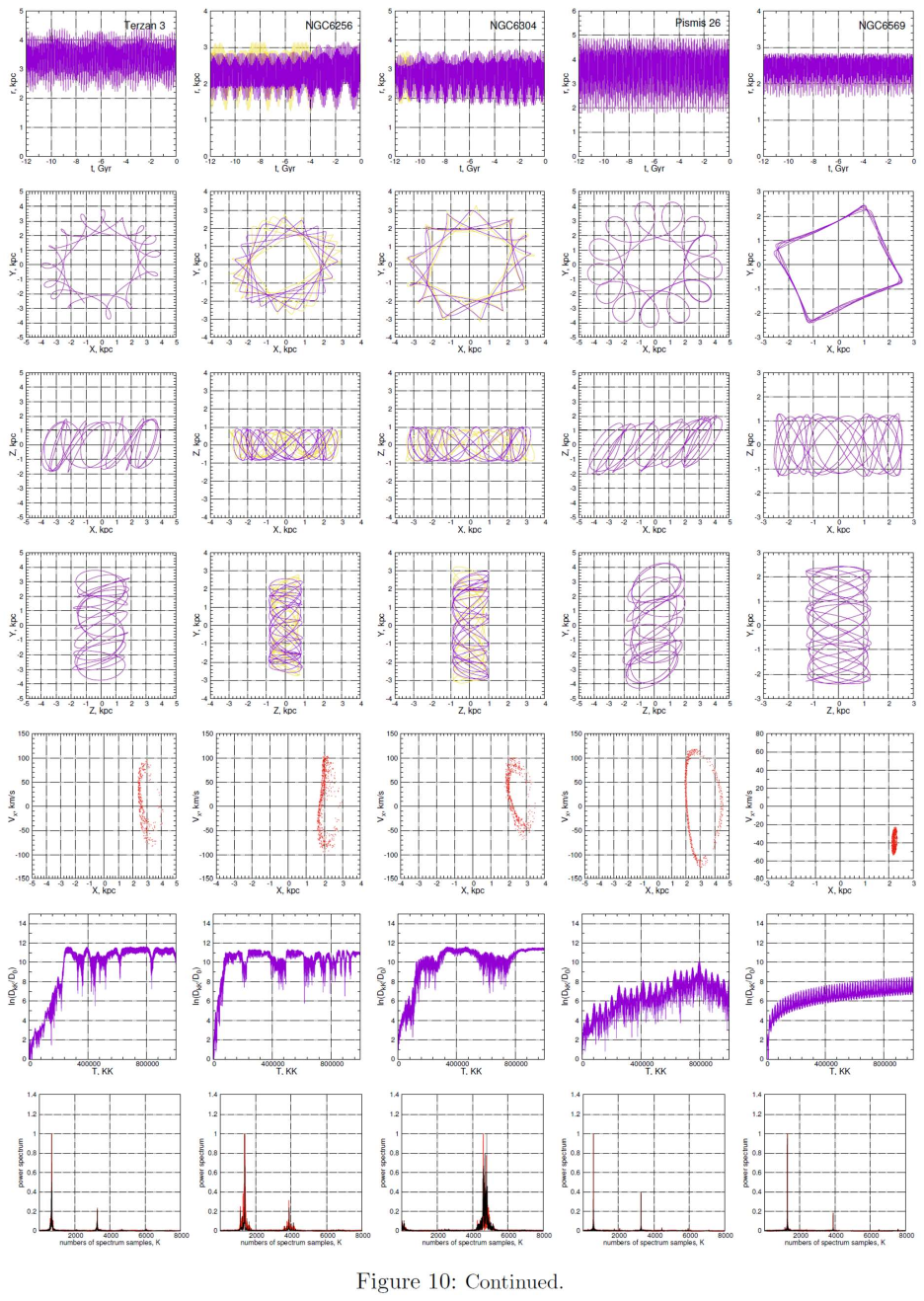}

\label{Lyap97}
\end{center}}
\end{figure*}

\newpage
\begin{figure*}
{\begin{center}

\includegraphics[width=1.0\textwidth,angle=0]{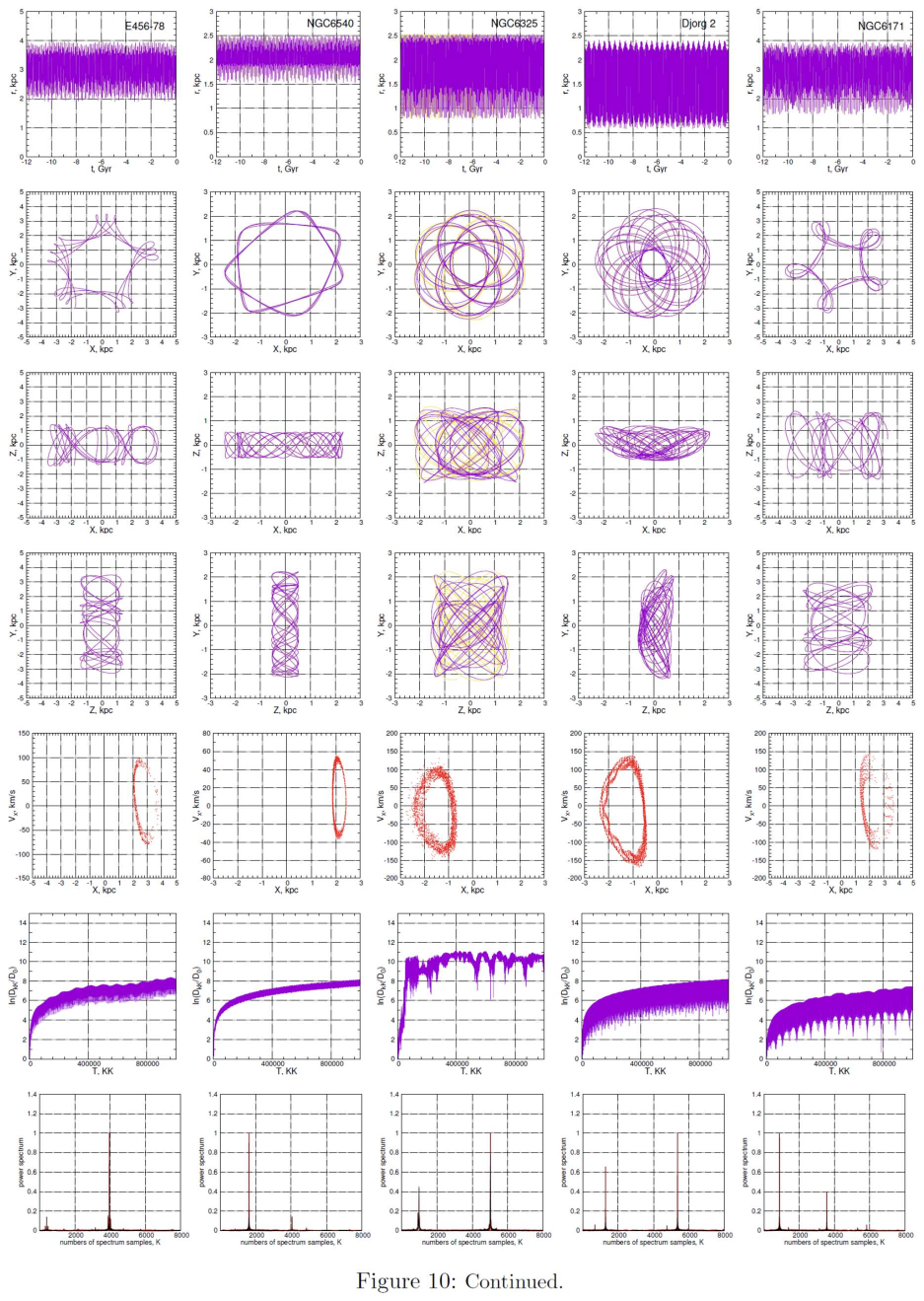}

\label{Lyap98}
\end{center}}
\end{figure*}

\newpage
\begin{figure*}
{\begin{center}

\includegraphics[width=1.0\textwidth,angle=0]{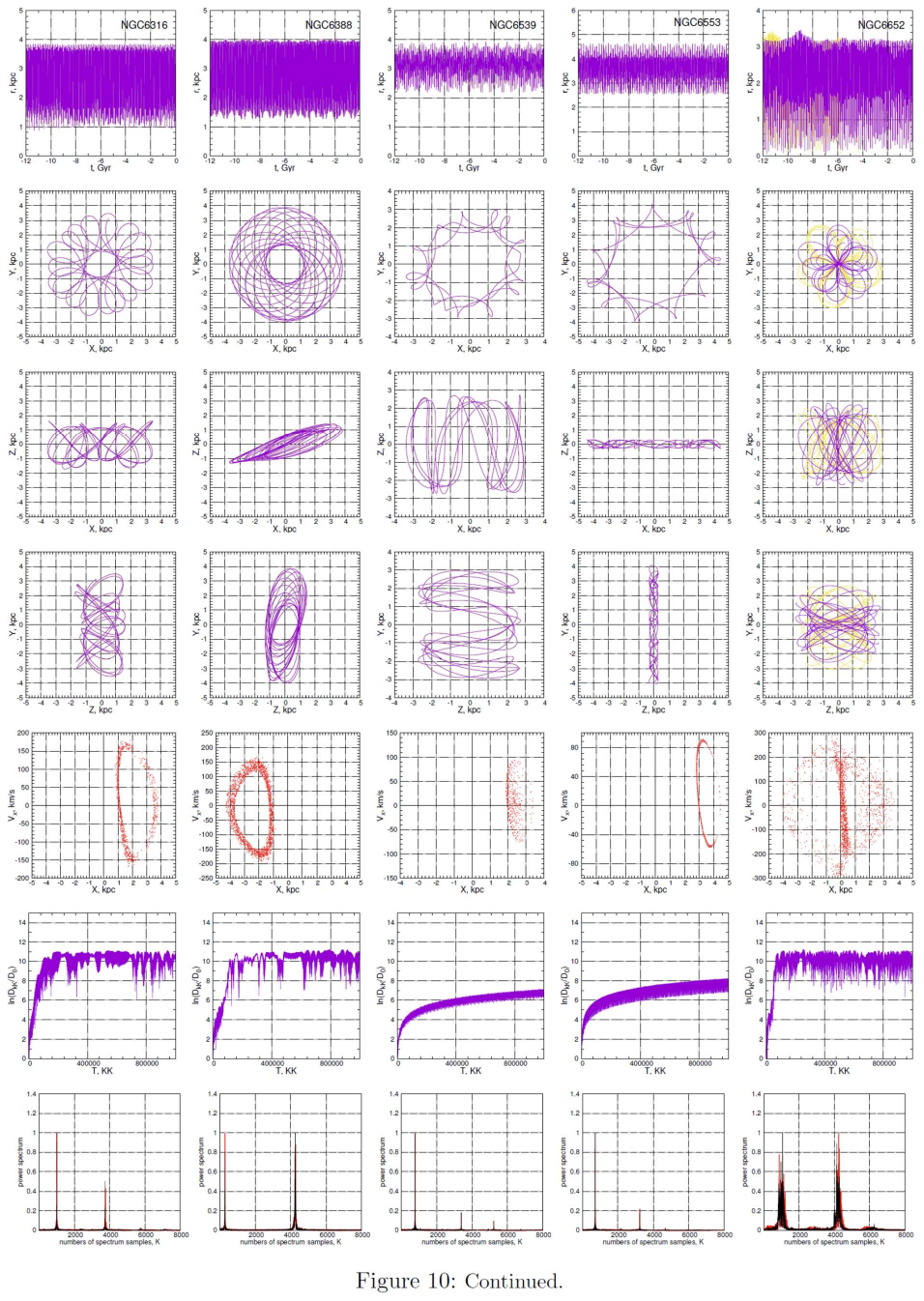}

\label{Lyap99}
\end{center}}
\end{figure*}


 {\begin{table*}[t]                                    
 \caption[]
 {\baselineskip=1.0ex
Summary table of signs of regularity (R) and chaoticity (C) of the 45 GC orbits.
  }
 \label{t:f}
 {\small\begin{center}\begin{tabular}{|r|l||c|c|c|c|c|c|c|}\hline
    &       & Proba-&MEGNO & HLE (1/Gyr)&Poincar$\acute{e}$ &Frequency&Visual&Probability \\
 N  &ID of  & bility&t=270 & t=120       & sections&drift    &assess- &P(R) by $"$voting$"$  \\
    & GC    & method&Gyr   & Gyr         &         &$\lg(\Delta f)$&ment&method \\
    &        &   (1)  & (2)     & (3)      & (4)   & (5)           & (6)   & (7)        \\\hline
 1  &NGC6144 & (C)      & 2.173  (R)& -0.002  (R)& (C)     &-2.08 (C)&  (C)    & 1/3 (C)\\\hline
 2  &E452-11 & (C)      & 0.752  (C)& 0.919   (C)& (C)     &-1.37 (C)&  (C)    &  0 (C) \\\hline
 3  &NGC6266 & (R)      & 1.976  (R)& -0.017  (R)& (R)     &-4.00 (R)&  (R)    &  1 (R) \\\hline
 4  &NGC6273 & (C)      & 1.494  (C)& 1.318   (C)& (C)     &-1.77 (C)&  (C)    &  0 (C) \\\hline
 5  &NGC6293 & (C)      & 0.934  (C)& 4.167   (C)& (C)     &-0.07 (C)&  (C)    &  0 (C) \\\hline
 6  &NGC6342 & (C)      & 0.769  (C)& 0.428   (C)& (C)     &-2.14 (C)&  (C)    &  0 (C) \\\hline
 7  &NGC6355 & (C)      & 0.509  (C)& 2.257   (C)& (C)     &-0.10 (C)&  (C)    &  0 (C) \\\hline
 8  &Terzan2 & (C)      & 0.627  (C)& 0.905   (C)& (C)     &-0.23 (C)&  (C)    &  0 (C) \\\hline
 9  &Terzan4 & (R)      & 1.993  (R)& -0.144  (R)& (R)     &-4.00 (R)&  (R)    &  1 (R) \\\hline
10  &BH229   & (C)      & 0.663  C& 2.220     (C)& (C)     &-1.81 (C)&  (C)    &  0 (C) \\\hline
11  &Liller1 & (R)      & 2.049  (R)& -0.037  (R)& (R)     &-4.00 (R)&  (R)    &  1 (R) \\\hline
12  &NGC6380 & (R)      & 2.182  (R)& 0.220   (C)& (R)     &-3.72 (R)&  (R)    &  5/6 (R)\\\hline
13  &Terzan1 & (R)      & 2.000  (R)& -0.029  (R)& (R)     &-4.00 (R)&  (R)    &  1 (R) \\\hline
14  &NGC6401 & (C)      & 0.622  (C)& 4.712   (C)& (C)     &-0.09 (C)&  (C)    &  0 (C) \\\hline
15  &Pal6    & (C)      & 0.502  (C)& 3.359   (C)& (C)     &-0.10 (C)&  (C)    &  0 (C) \\\hline
16  &Terzan5 & (R)      & 2.023  (R)& 0.041   (C)& (R)     &-4.00 (R)&  (R)    &  5/6 (R)\\\hline
17  &NGC6440 & (C)      & 1.901  (R)& 0.572   (C)& (R)     &-2.26 (C)&  (C)    &  1/3 (C) \\\hline
18  &Terzan6 & (R)      & 1.996  (R)& -0.055  (R)& (R)     &-4.00 (R)&  (R)    &  1 (R) \\\hline
19  &NGC6453 & (C)      & 1.178  (C)& 1.998   (C)& (C)     &-1.92 (C)&  (C)    &  0 (C) \\\hline
20  &Terzan9 & (R)      & 2.358  (R)& -0.056  (R)& (R)     &-3.86 (R)&  (R)    &  1 (R) \\\hline
21  &NGC6522 & (R)      & 1.996  (R)& -0.020  (R)& (R)     &-4.00 (R)&  (R)    &  1 (R) \\\hline
22  &NGC6528 & (R)      & 2.008  (R)& -0.036  (R)& (R)     &-4.00 (R)&  (R)    &  1 (R) \\\hline
23  &NGC6558 & (C)      & 0.819  (C)& 1.364   (C)& (C)     &-1.03 (C)&  (C)    &  0 (C) \\\hline
24  &NGC6624 & (R)      & 1.847  (R)& -0.040  (R)& (R)     &-4.00 (R)&  (R)    &  1 (R) \\\hline
25  &NGC6626 & (C)      & 1.194  (C)& 0.093   (C)& (C)     &-1.78 (C)&  (C)    &  0 (C) \\\hline
26  &NGC6638 & (C)      & 0.533  (C)& 2.411   (C)& (C)     &-0.16 (C)&  (C)    &  0 (C) \\\hline
27  &NGC6637 & (R)      & 1.988  (R)& -0.012  (R)& (R)     &-4.00 (R)&  (R)    &  1 (R) \\\hline
28  &NGC6642 & (C)      & 0.681  (C)& 2.451   (C)& (C)     &-1.01 (C)&  (C)    &  0 (C) \\\hline
29  &NGC6717 & (R)      & 2.044  (R)& -0.001  (R)& (R)     &-4.00 (R)&  (R)    &  1 (R) \\\hline
30  &NGC6723 & (R)      & 2.252  (R)& 0.064   (C)& (R)     &-4.00 (R)&  (R)    &  5/6 (R)\\\hline
31  &Terzan3 & (R)      & 3.495  (R)& -0.000  (R)& (R)     &-1.89 (R)&  (R)    &  1 (R) \\\hline
32  &NGC6256 & (C)      & 0.893  (C)& -0.000  (R)& (C)     &-1.93 (C)&  (C)    &  1/6 (C)\\\hline
33  &NGC6304 & (R)      & 1.753  (R)& -0.000  (R)& (C)     &-1.38 (C)&  (C)    &  1/2 (C)\\\hline
34  &Pismi26 & (R)      & 1.941  (R)& -0.000  (R)& (R)     &-4.00 (R)&  (R)    &  1 (R) \\\hline
35  &NGC6569 & (R)      & 1.957  (R)& -0.000  (R)& (R)     &-4.00 (R)&  (R)    &  1 (R) \\\hline
36  &E456-78 & (R)      & 1.983  (R)& -0.000  (R)& (R)     &-3.59 (R)&  (R)    &  1 (R) \\\hline
37  &NGC6540 & (R)      & 1.999  (R)& -0.000  (R)& (R)     &-4.00 (R)&  (R)    &  1 (R) \\\hline
38  &NGC6325 & (C)      & 1.216  (C)& -0.000  (R)& (C)     &-3.22 (R)&  (C)    &  1/3 (C)\\\hline
39  &Djorg2  & (R)      & 2.320  (R)& -0.050  (R)& (R)     &-4.00 (R)&  (R)    &  1 (R) \\\hline
40  &NGC6171 & (R)      & 2.015  (R)& -0.000  (R)& (R)     &-4.00 (R)&  (R)    &  1 (R) \\\hline
41  &NGC6316 & (R)      & 2.289  (R)& 0.251   (C)& (R)     &-1.96 (R)&  (R)    &  5/6 (R)\\\hline
42  &NGC6388 & (R)      & 2.450  (R)& 0.271   (C)& (C)     &-0.03 (C)&  (R)    &  1/2 (C)\\\hline
43  &NGC6539 & (R)      & 1.993  (R)& -0.000  (R)& (R)     &-4.00 (R)&  (R)    &  1 (R) \\\hline
44  &NGC6553 & (R)      & 1.899  (R)& -0.000  (R)& (R)     &-4.00 (R)&  (R)    &  1 (R) \\\hline
45  &NGC6652 & (C)      & 1.121  (C)& 3.269   (C)& (C)     &-0.12 (C)&  (C)    &  0 (C) \\\hline
 \end{tabular}\end{center}}\end{table*}}

 {\begin{table*}[t]                                    
 \caption[]
 {\baselineskip=1.0ex
Correlation matrix 1
  }
 \label{t:f2}
 {\begin{center}\begin{tabular}{|r|c|c|c|}\hline
Method&  (2)   & (3)   & (5)    \\\hline
(2)  &   1.00 & 0.70  & 0.65   \\\hline
(3)  &   0.70 & 1.00  & 0.76   \\\hline
(5)  &   0.65 & 0.76  & 1.00   \\\hline

\end{tabular}\end{center}}\end{table*}}

 {\begin{table*}[t]                                    
 \caption[]
 {\baselineskip=1.0ex
Correlation matrix 2
  }
 \label{t:f2}
 {\begin{center}
 \begin{tabular}{|r|c|c|c|c|c|c|c|}\hline
Method&  (1)     & (2)  & (3)  & (4)     & (5)       & (6)       & (7)   \\\hline
(1)  &  1.00    & 0.95 & 0.60 & 0.91    & 0.86      &  0.96     & 0.96  \\\hline
(2)  &  0.95    & 1.00 & 0.65 & 0.87    & 0.82      &  0.91     & 0.95  \\\hline
(3)  &  0.60    & 0.65 & 1.00 & 0.60    & 0.64      &  0.55     & 0.74  \\\hline
(4)  &  0.91    & 0.87 & 0.60 & 1.00    & 0.96      &  0.96     & 0.96  \\\hline
(5)  &  0.86    & 0.82 & 0.64 & 0.96    & 1.00      &  0.91     & 0.94  \\\hline
(6)  &  0.96    & 0.91 & 0.55 & 0.96    & 0.91      &  1.00     & 0.96  \\\hline
(7)  &  0.96    & 0.95 & 0.74 & 0.96    & 0.94      &  0.96     & 1.00  \\\hline
\end{tabular}\end{center}}
\end{table*}}


\begin{thebibliography}{25}
\providecommand{\natexlab}[1]{#1}
\bibitem
1~A. T. Bajkova and V. V. Bobylev. Izvestiya Glavnoi astronomicheskoi observatorii v
Pulkove {\bf 227}, 15 (2022) DOI:10.31725/0367-7966-2022-227-3, arXiv: 2212.00739.
\bibitem
2~A. T. Bajkova, A. A. Smirnov, and V. V. Bobylev, Izvestiya Glavnoi astronomicheskoi observatorii v Pulkove {\bf 228}, 1 (2023) DOI:10.31725/0367-7966-2023-228-1, arXiv: 2305.05012.
\bibitem
3~A. T. Bajkova, A. A. Smirnov, and V. V. Bobylev, Izvestiya Glavnoi astronomicheskoi observatorii v Pulkove {\bf 229}, 1 (2023) DOI:10.31725/0367-7966-2023-229-1.
\bibitem
4~A. A. Smirnov, A. T. Bajkova, and V. V. Bobylev, Izvestiya Glavnoi astronomicheskoi observatorii v Pulkove {\bf 228}, 157 (2023) DOI:10.31725/0367-7966-2023-228-12.
\bibitem
5~A. T. Bajkova, A. A. Smirnov, and V. V. Bobylev, Astrophysical Bulletin {\bf 78}, Issue, 4, 499 (2023), arXiv: 2311.14789.
\bibitem
6~A. A. Smirnov, A. T. Bajkova, V. V. Bobylev, Monthly Notices of the Royal Astronomical Society {\bf 528}, Issue 2, 1422 (2024), arXiv: 2310.18172.
\bibitem
7~E. Vasiliev, H. Baumgardt, Monthly Notices of the Royal Astronomical Society {\bf 505}, Issue 4, 5978 (2021), arXiv: 2102.09568.
\bibitem
8~H. Baumgardt and E. Vasiliev, Monthly Notices of the Royal Astronomical Society {\bf 505}, Issue 4, 5957 (2021), arXiv: 2105.09526.
\bibitem
9~J. Palou$\breve{s}$, B. Jungwiert, J. Kopeck$\acute{y}$, Astronomy and Astrophysics {\bf 274}, 189 (1993).
\bibitem
1~J. L. Sanders, L. Smith, N. W. Evans, P. Lucas, Monthly Notices of the Royal Astronomical Society  {\bf 487}, Issue 4, 5188 (2019), arXiv: 1903.02008.
\bibitem
1~R. E. G. Machado, T. Manos, Monthly Notices of the Royal Astronomical Society {\bf 458}, Issue 4, 3578 (2016), arXiv: 1603.02294.
\bibitem
1~A. V. Mel'nikov, Solar System Research {\bf 52} (5), 417 (2018).
\bibitem
1~C. D. Murray, S. F. Dermott. {\it Solar System Dynamics}. Cambridge University Press, 592 p. (2000).
\bibitem
1~A. Morbidelli. {\it Modern Celestial Mechanics. Aspects of Solar System Dynamics.} London and New York, 355 p. (2011).
\bibitem
1~S. Breiter, B. Melendo, P. Bartczak, I. Wytrzyszczak, Astronomy and Astrophysics, {\bf 437}, Issue 2, 753 (2005).
\bibitem
1~N. Nieuwmunster, M. Schultheis, M. Sormani, F. Fragkoudi, F. Nogueras-Lara, R. Schodel, P. McMillan, L. Smith, J. Sanders, Astronomy and Astrophysics, {\bf 685}, id.A93, 19 pp. (2024), arXiv: 2403.00761.
\bibitem
1~M. Valluri, V. P. Debattista, T. Quinn, B. Moore, Monthly Notices of the Royal Astronomical Society {\bf 403}, Issue 1, 525 (2010), arXiv: 0906.4784.
\end{thebibliography}
\end{document}